\documentclass[a4paper,10pt, final]{IEEEtran}

\usepackage[english]{babel}
\usepackage{hyperref}
\usepackage[T1]{fontenc}
\usepackage[cmex10]{amsmath}
\usepackage{amsfonts,amssymb,bm,amsthm}
\usepackage{graphicx}
\usepackage[nolist]{acronym}
\usepackage[margin=.5in,footskip=0.1in]{geometry}
\usepackage{graphics, graphicx}
\usepackage{amssymb,amsmath}
\usepackage{amsmath}
\usepackage{threeparttable}  
\usepackage{url}   
\usepackage[ruled,vlined,linesnumbered]{algorithm2e}
\usepackage{algpseudocode}   
\usepackage{cite}
\usepackage{multirow}
\usepackage{lineno}
\usepackage[normalem]{ulem}
\usepackage{indentfirst}
\usepackage{caption}
\usepackage{subcaption}
\usepackage{wrapfig}
\usepackage{fixltx2e}
\usepackage{xfrac}
\usepackage{cancel}
\usepackage{cite}
\usepackage{booktabs}
\usepackage{schemabloc}
\usetikzlibrary{arrows, calc, positioning}
\usepackage{standalone}
\usepackage{geometry}
\usepackage{multicol}
\usepackage{comment}
\usepackage{hyperref}

\usepackage{tikz}
\definecolor{gold}{rgb}{0.85,.66,0}
\definecolor{viol}{rgb}{.68, .75, 1}

\DeclareMathAlphabet\mathbfcal{OMS}{cmsy}{b}{n}

\begin{document}
\begin{acronym}[p1]

\acro{TAS}{Transmitter Antenna Selection}
\acro{THz}{Terahertz}
\acro{MIMO}{Multiple-Input Multiple-Output}
\acro{mMIMO}{massive Multiple-Input Multiple-Output}
\acro{MISO}{Multiple-Input Single-Output}
\acro{NOMA}{Non-Orthogonal Multiple Access}
\acro{OMA}{Orthogonal Multiple Access}
\acro{5G}{Fifth Generation}
\acro{BS}{Base Station}
\acro{NOMA-MIMO}{Non-Orthogonal Multiple Access Multiple-Input Multiple-Output}
\acro{ML}{Machine Learning}
\acro{CNN-TAS}{Convolutional Neural Network-based Transmitter Antenna Selection}
\acro{SE}{Spectral Efficiency}
\acro{EPA}{Equal Power Allocation}
\acro{PICPA}{Proportional to the Inverse of the Channel Power Allocation}
\acro{EE}{Energy Efficiency}
\acro{TDMA}{Time Division Multiple Access}
\acro{FDMA}{Frequency Division Multiple Access}
\acro{CDMA}{Code Division Multiple Access}
\acro{RF}{Radio Frequency}
\acro{AS}{Antenna Selection}
\acro{GA}{Genetic Algorithm}
\acro{IoT}{Internet of Things}
\acro{RM}{Resource Management}
\acro{CNN}{Convolutional Neural Network}
\acro{OFDM}{Orthogonal Frequency Division Multiplexing}
\acro{FNN}{Fully-connected Neural Networks}
\acro{MMSE}{Minimum Mean-Squared Error}
\acro{SL}{Supervised Learning}
\acro{DL}{Downlink}
\acro{CSI}{Channel State Information}
\acro{ZFBF}{Zero-Forcing Beamforming}
\acro{EPA}{Equal Power Allocation}
\acro{UE}{User Equipment}
\acro{SIC}{Successive Interference Cancellation}
\acro{SR}{Sum-Rate}
\acro{AIA-AS}{max-min-max AS}
\acro{SINR}{Signal to Interference plus Noise Ratio}
\acro{NN}{Neural Network}
\acro{NLOS}{Non-Line-Of-Sight}
\acro{GD}{Gradient Descend}
\acro{ES}{Exhaustive Search}
\acro{QoS}{Quality of Service}
\acro{R-TAS}{random-Transmitter Antenna Selection}
\acro{HRPN}{Hyper Region Proposal Network}
\acro{GA}{Genetic Algorithm}
\acro{TDD}{Time Division Duplex}
\acro{ZF}{Zero Forcing}
\acro{WF}{Water-Filling}
\acro{PA}{Power Allocations}
\end{acronym}

\title{\textbf{Massive MIMO and NOMA Bits-per-Antenna Efficiency under Power Allocation Policies}}
\author{\textbf{Thiago A. Bruza Alves}, \textbf{Taufik Abrão}\\
State University of Londrina (UEL), Department of Electrical Engineering, Londrina-PR, Brazil.
\thanks{This work was partly supported by The National Council for Scientific and Technological Development (CNPq) of Brazil under Grants 310681/2019-7, partly by the CAPES- Brazil - Finance Code 001,  and the Londrina State University - Paraná State Government (UEL).}
\thanks{T. A. Bruza Alves and Taufik Abrão are with the State University of Londrina (UEL), Department of Electrical Engineering, Londrina-PR, E-mails: thiagobruza@hotmail.com, taufik@uel.br}}

\maketitle

\begin{abstract}
A comparative resource allocation analysis in terms of received bits-per-antenna spectral efficiency (SE) and energy efficiency (EE)  in downlink (DL) single-cell massive multiple-input multiple-output (mMIMO) and non-orthogonal multiple access (NOMA) systems considering a BS equipped with many ($M$) antennas, while  $K$ devices {operate} with a single-antenna, and {the loading of devices} $\rho = \frac{K}{M}$ ranging in $0<\rho\leq 2$ is carried out under three different \ac{PA} strategies: the inverse of the channel power allocation (PICPA), a  modified water-filling ($\Delta$-WF) allocation method, and the equal power allocation (EPA) reference method.  Since the two devices per cluster are overlapped in the power domain in the NOMA system, the channel matrix requires transformation to perform the zero-forcing (ZF) precoding adopted in mMIMO. Hence, NOMA operating under many antennas can favor a group of devices with higher array gain, overcoming the mMIMO and operating conveniently in the higher loading range $0.6<\rho<2.0$. In such a scenario, a more realistic and helpful metric consists in {evaluating} the area under SE and EE curves, by measuring the bit-per-antenna and bit-per-antenna-per-watt efficiency, respectively. Our numerical results confirm a superiority of NOMA w.r.t. mMIMO of an order of 3x for the SE-area and 2x for the EE-area metric.
\end{abstract}
	
\smallskip


\section{Introduction}
The {beyond} \ac{5G} of wireless communication systems must {allow} ultra-dense connections with vastly heterogeneous requirements. The challenges in networks persist, including the \ac{SE} and the \ac{EE} joint improvement, the increase in the SE-EE trade-off, and \ac{QoS}, always aiming to meet the growing number of devices connected to the network.  Among the proposals to solve these {challenges}, the \ac{mMIMO} system is the primary proposed system that allows the increase of the link capacity, exploring the propagation of multiple paths with the use of a large number of antennas at the \ac{BS} \cite{Larsson2014,Bjornson2014a}.  Another relevant enabling technology is the \ac{NOMA}, which explores the power domain as an alternative way in terms of multiple access technology, helping to mitigate the spectrum exhaustion problem and serving more than one device per resource block \cite{Ding2016b}.

Although in many works \ac{mMIMO} is classified as an orthogonal technique, allocating the signal from devices in the same resource block, possible by spatial diversity, allows us to classify it as a non-orthogonal technique too \cite{Senel2019}.  There is a vast literature demonstrating the superior performance of the Spectral Efficiency of \ac{NOMA} when compared to \ac{OMA} techniques \cite{Anwar2019}.  Previous aims to improve the communication system performance by combining MIMO  ({with a small number of antennas $M$}) and NOMA have been discussed in \cite{Ali2017,Zeng2017b,Islam2018, Chen2018}.

Studies comparing \ac{NOMA} and \ac{mMIMO} {in a single cell are proposed in \cite{Cheng2018, Dai2019, Senel2019}}. The acquisition of \ac{CSI} through pilot acquisition to \ac{NOMA} system {is proposed in \cite{Cheng2018}}. In \cite{Dai2019}, the application of NOMA in the mMIMO scheme is proposed, and better results are achieved in the proposed comparative. Moreover, in \cite{Senel2019} is analyzed the performance of NOMA and mMIMO in line of sight and non-line of sight.

The canonical \ac{mMIMO} refers to the systems with \ac{BS}s formed by a large number of antennas $M$ when compared to the number of actives devices, $K$, succinctly $M \gg K$ is considered a \ac{mMIMO} setup. The typical \ac{NOMA} improves the \ac{SE} by superposing the signals of the selected devices to form a cluster in {the power domain}, multiplexing it over the same signal and served by the same beamforming. {Nonetheless, the success} of \ac{NOMA} depends on the \ac{SIC}. 

Power-domain NOMA can be a candidate technology in dense networks \cite{Makki2020}. To improve performance and minimize the impact to assume the perfect \ac{SIC} \cite{Usman2017}, devices are divided {into} two groups. After grouping in pairs and forming a cluster, each pair forms a cluster with a high difference between channel conditions. The device with a higher channel condition can decode the signal sent to the device with the lower channel condition. The interference can thus be eliminated by \ac{SIC}. The use of \ac{NOMA} in BS equipped with a large number of antennas was investigated in terms of SE {\cite{Senel2019,Cheng2018}} we propose in a similar configuration system increasing the loading up to 2 times the number of antennas in BS and analyze the SE, EE, and SE-EE trade-off.

The \ac{EE} metric is a popular figure of merit employed to analyze the balance between power consumption and data rate. The \ac{EE} is the ratio between the effectively transmitted data rate and the total power expended during the transmission process, including instantaneous and static components. With the \ac{EE} metric, it is possible to evaluate the efficiency with which a system uses the limited energy resource to communicate data and optimize this ratio. Can show the tendency of energy consumption in the case of seeking justice among devices.

{The \ac{ZF} is simple and popular alternative interference suppression} beamforming under perfect \ac{CSI} condition and achieving a satisfactory condition in real situations when imperfect \ac{CSI}, in this work, we adopt perfect \ac{CSI}, for that the pilots are needed. {The adoption of \ac{NOMA} system with a large number of antennas requires a defined equivalent channel to be deployed for interference mitigation; and according to the \ac{NOMA} principle, makes the equivalent channel matrix smaller than the original one due to the exploration of power domain in NOMA.} 
 
{Various transmission topologies already deal with the \ac{EE} problem in \ac{mMIMO}, finding the optimal number of antennas, number of devices in a cell, and the maximal \ac{EE} \cite{Bjornson2014a,Bjornson2019}. The EE analysis in the NOMA system is carried out} in \cite{Zhang2017a}, and its superiority {is demonstrated} when compared with conventional orthogonal multiple access (OMA) systems. Recent researches seek to improve the \ac{NOMA} performance, {\it e.g.} in \cite{Mounchili2020}, the minimum pairing distance is defined and compared to the \ac{OMA}, while in \cite{Rezaei2020} it's presented {a comparison} between \ac{OMA} and cell-free system equipped with \ac{mMIMO}-\ac{NOMA}. An \ac{EE} analysis in \ac{THz}-\ac{NOMA}-\ac{MIMO} was proposed in \cite{Zhang2020}. Still, the number of active devices is much smaller than the number of antennas in the BS, and \cite{Maraqa2020} is a survey about Power Domain \ac{NOMA} and makes clear the vacuum of \ac{EE} analysis and comparison between \ac{NOMA} with many antennas and \ac{mMIMO}.

Recent works propose the deployment of \ac{NOMA} combined with other techniques {a more} effective transmission scheme; {\it e.g.}, in \cite{Rajoria2021} \ac{NOMA} and \ac{mMIMO} are jointly considered in a two-tier network for accommodating colossal traffic. Furthermore, in \cite{Kim2021}, authors apply NOMA in \textit{Distributed Antenna Systems} (DAS), aiming to achieve better performance when compared to the conventional NOMA or DAS technique alone. While \cite{Budhiraja2021} shows an in-depth survey of {the state-of-the-art} of power-domain NOMA variants; moreover, several open issues and research challenges of NOMA-based applications are systematized. {The NOMA system presents drawbacks, such as hardware (including SIC) complexity, channel feedback, receiver design, and careful power and pilot allocation strategies \cite{Makki2020, Maraqa2020,Zafari2019}.}

This work focus on {revealing the advantages} of applying the \ac{mMIMO} scheme {\it versus} \ac{NOMA} scheme with a massive number of BS antennas, and varying {the loading of devices}, {\it i.e.,} the ratio of the number of mobile devices to the number of BS antennas, $\rho = \frac{K}{M}$, while we change the \ac{PA} strategy. Besides, we adopt a realistic model for the system's power consumption as in \cite{Bjornson2014a} but adapted to our needs, aiming at providing a suitable analysis of the system {resource allocation}.

\vspace{2mm}
\noindent \textbf{\textit{Contributions:}} the contributions of this work are fourfold. {\bf a}) an extensive and comparative analysis on the spectral efficiency (SE) performance of mMIMO system against NOMA system, varying the system loading under specific (three different) power allocation methods and making use of the area under the SE ($\mathcal{S}^{\textsc{syst}}$) curve of the system as an effective, useful and fair metric of performance and efficiency; {\bf b}) we develop an energy efficiency (EE) analysis using a detailed model of energy consumption, with fixed and variable terms related to circuitry power consumption with number of antennas and devices, respectively, providing an extensive and comparative analysis on both the NOMA and mMIMO systems under realistic operation scenarios and making use of the area under the EE ($\mathcal{E}^{\textsc{syst}}$) curve of system; {\bf c}) an analysis on the SE-EE trade-off is developed considering a wide range of {loading of devices}, verifying the fairness between devices; {\bf d}) finally, under mild conditions, we provide evidences for the NOMA’s ability to serve a greater number of devices than mMIMO system. 

The remainder of the paper is organized as follows.  Section \ref{sec:model} describes the system models for NOMA and mMIMO adopted in this work. In Section \ref{sec:EE-SE} we present the proposed EE-SE formulation for NOMA and massive MIMO systems. Numerical results are analyzed in \ref{sec:results}. Section \ref{sec:concl} concludes the paper.
 
\vspace{2mm}
\noindent {\it Notation}. In this work, boldface lower case and upper case characters denote vectors and matrices, respectively. The operator $(x)^+ = \max(0,x)$. The operators  $[\cdot  ]^{\text{T}}$, $\mathbb{E}[\cdot]$ and $|\cdot|$ denote transpose,  expectation and cardinality, respectively.  A random vector ${\bf x} \sim \mathcal{CN} \left \{0,\mathbf{I}_m  \right \}$ is circularly symmetric Gaussian distributed with mean $0$ and covariance matrix $\mathbf{I}_m$. $\mathbf{I}_m$ is $m \times m$ identity matrix.

\section{System Models}\label{sec:model}
Let us consider a multi-user single-cell \ac{DL} transmission operating in a \ac{TDD} with $K$ single-antenna actives devices, communicating with one \ac{BS}, which is equipped with $M$ transmit antennas in \ac{NLOS}.  The set $\mathcal{K}$ is formed by $K$ devices, {these devices} are randomly distributed in a radius disk $d_{\max}$, the disk area is formed by two sub-disk with the same number of devices in each sub-area; both subsets are identified as $\mathcal{K}_H$ and $\mathcal{K}_L$.  In the first subset, $\mathcal{K}_H$ represents devices' indexes having the higher channel coefficient and sort in descending order, while the other subset $\mathcal{K}_L$ are formed by the devices with lower channel coefficient and sort in ascending order; the indexes $k \in \mathcal{K}_H$ and $k \in \mathcal{K}_L$ such that:
\begin{equation} \label{eq:sets}
\begin{split}
&\mathcal{K} = \mathcal{K}_H \cup \mathcal{K}_L,\quad \text{where}\,\,\, \\
&\mathcal{K}_H=\{1,...,K/2\} \,\,\, \text{and}\,\,\, \mathcal{K}_L = \{K/2 + 1,...,K\}. 
\end{split}
\end{equation} 
The channel vector modeling of device $k$ can be described liked as:
\begin{equation} \label{eq:channel}
\mathbf{h}_{k} = \sqrt{\beta_{k}} \mathbf{h}_{k}', \quad k = 1,..., K,
\end{equation}
where $\beta_{k}$ is the large-scale fading coefficient and satisfy
\begin{equation}
\beta_{j} > \beta_{i}, \quad \forall j \in \mathcal{K}_H, \,\,\,  \forall i \in \mathcal{K}_L.
\end{equation}
Herein, the pathloss model in [dB] is defined as:
\begin{equation}
\beta_k = \beta_0 + 10\cdot \xi \cdot \text{log}_{10}(d_k),
\end{equation}
where $d_k$ is the distance of user $k$ to \ac{BS}, $\xi$ is the pathloss coefficient, and $\beta_0$ is the attenuation at the distance of reference. 

In each coherence interval, $\mathbf{h}'_{k}$ in (\ref{eq:channel}) for device $k$ is an independent random small-scale fading realization from an independent Rayleigh fading distribution, $\mathbf{h}'_{k} \sim \mathcal{CN}(0,\mathbf{I}_M), k = 1,...,K$ . The transmitted signal $\mathbf{x}_{k} \in \mathbb{C}^{M}$ is the beamformed data symbol of device $k$:
\begin{equation}
\mathbf{x}_{k} = \mathbf{g}_{k} \sqrt{p_{k}} s_{k},
\end{equation}
where $\mathbf{g}_{k}$ is a normalized beamforming vector, $p_{k}$ normalized transmission power and $s_{k} \sim \mathcal{CN}(0,1)$ is the data symbol of device $k$, and period $T_\textsc{s}$. 
The signal received at the $k$-th device:
\begin{align}
y_{k} &= \sum^{K}_{k'=1} \mathbf{h}_{k}^{\text{T}}\mathbf{x}_{k'} + n_{k}, \notag\\
& = \sqrt{\beta_{k}} \mathbf{h'}_{k}^{\text{T}} \sum^{K}_{k'=1}  \mathbf{g}_{k'} \sqrt{p_{k'}} s_{k'} + n_{k},\\
& = \sqrt{\beta_{k}p_{k}} \mathbf{h'}_{k}^{\text{T}} \mathbf{g}_{k} s_{k} +  \sqrt{\beta_{k}} \mathbf{h'}_{k}^{\text{T}}  \sum^{K}_{\substack{k'\neq k \\ k'=1}}  \mathbf{g}_{k'} \sqrt{p_{k'}} s_{k'} + n_{k},\notag
\end{align}	
where $n_{k} \sim \mathcal{CN}(0,1)$ is the additive noise. Notice that this modeling applies to both \ac{NOMA} and \ac{mMIMO} systems, but beamforming is selected differently, and this topic will be addressed in the next sections. 

\subsection{{Prior Actions}} \label{sec:Previous}
Because the BS needs to know {\it a priori} crucial information related to the channel and devices distributed in the cell, including device location, rate demanded, and channel coefficient, such required {\it a priori} information may differ depending on the multiple access scheme considered \cite{Makki2020}.

The option for the mMIMO and NOMA systems was carried out with the guarantee that the needs of the devices would be met, and this initial step was carried out successfully. Subsection \ref{sec:Previous2} briefly discusses the preliminary information required to proceed with different Power Allocation (PA) procedures in both mMIMO and NOMA systems.

\subsection{Pilot Overhead for Channel State Information} \label{sec:model_pilot}
\begin{figure}[htbp!]
\centering
\includegraphics[trim=.15cm 9.2cm .15cm 10cm, clip, width=.75\linewidth]{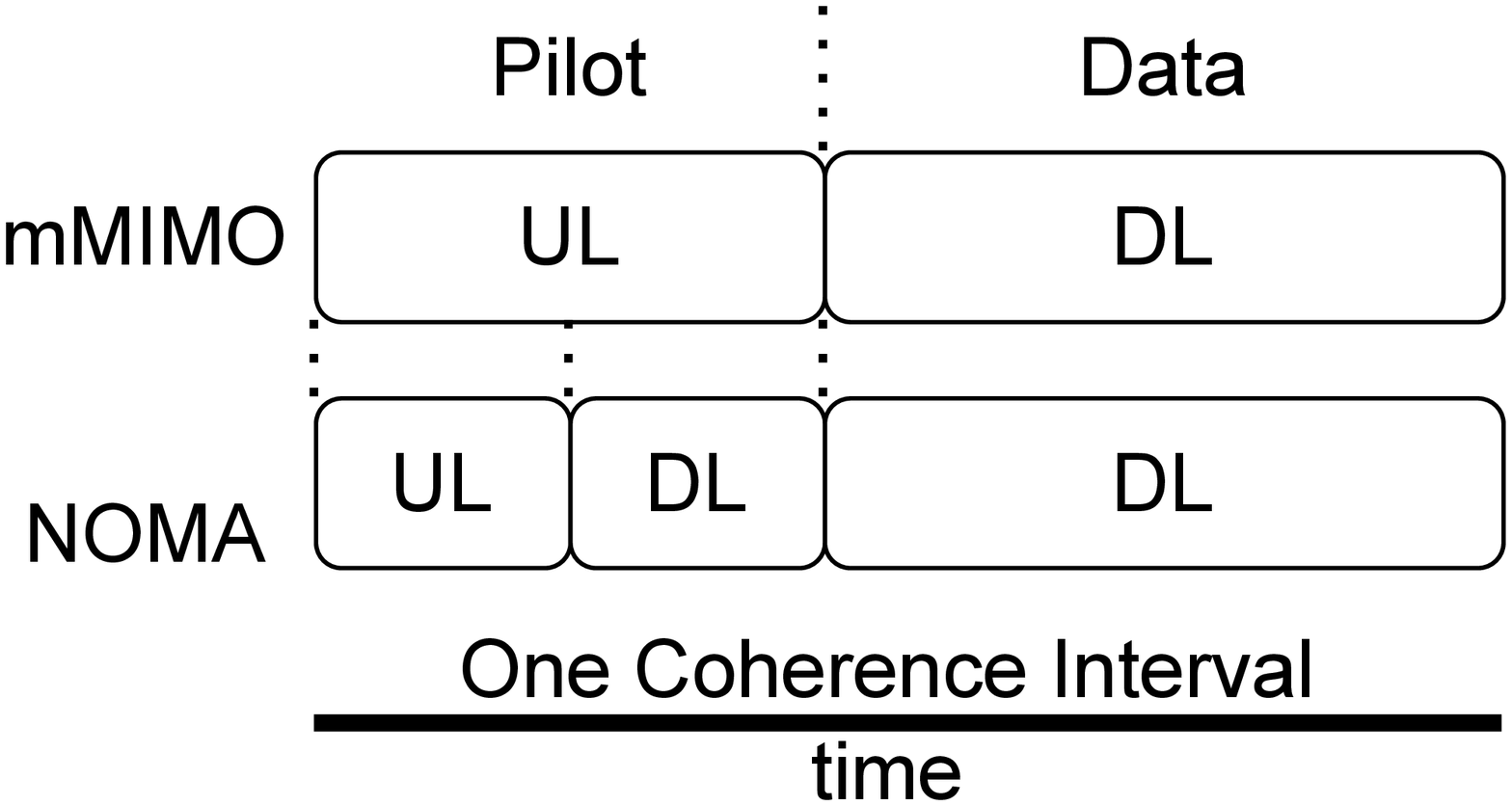}
\caption{Coherence time interval structure: the training and data transmission structure for mMIMO and NOMA schemes under TDD NLOS setup.}
\label{fig:coherence}
\end{figure}

Fig. \ref{fig:coherence} compares the pilot-data transmission structure along the one-channel coherence interval for both mMIMO and NOMA systems considered. Notice that $T_\textsc{s}$ is the time required to transmit a data symbol (Data). Moreover,  the channel coherence time interval  T is assumed to be a multiple of the Data symbol period,  $\textrm{T} = \iota \cdot T_\textsc{s}$. The power allocated to each pilot in the training step is enough. In contrast, the number of pilots, and the dedicated portion of coherence interval for data transmission are assumed to be the same in both systems.

Notice that in NOMA transmission, the pilot transmission step is split into two portions, half for the UL transmission pilots and receive the DL pilot confirmation; this happens because to perform SIC, the cell-center devices need to learn the effective channels that are established by the beamforming. Additionally, the beamforming vectors are based on cell-center devices, producing limited rates achieved in cell-edge devices. On the other hand, in the mMIMO scheme, a significant advantage is that there is no need for DL pilots since the effective channels created by the beamforming are highly predictable, {\it i.e.} nearly deterministic gain and phase due to channel hardening effect \cite{Senel2019}.

\vspace{2mm}
\noindent\textit{Assumption 1:} In the \ac{NOMA} system, the power allocated to each downlink pilot is sufficient to reach the destination device.\\

\subsection{Beamforming for NOMA and mMIMO systems} \label{sec:beam}
{At the \ac{mMIMO} system,} each device is served by a single beamforming vector. {The ZF technique is a popular interference-suppressing beamforming scheme in the mMIMO system since it eliminates all inter-user interference using individual beamforming for each device, while the favorable propagation facilitates such interference suppressing in massive MIMO configurations. Besides, to perform ZF precoding in NOMA system, it is essential to understand the NOMA \textit{user-pairing} concept}.

\vspace{2mm}
\noindent\textbf{\textit{{User-pairing:}}} {Inherent to the NOMA system, user clustering can be performed in several ways after the \textit{user-sorting} and the user classification in center-users and edge-users subsets. Because we know that the SE of NOMA is directly proportional to the difference between the pathloss of the users, a natural choice consists in pairing users with as higher as possible pathloss differences \cite{Ali2017}:
\begin{equation}\label{eq:DeltaBeta}
\Delta\beta_k = \beta_k - \beta_{K+1-k},
\end{equation}
forming the cluster $k$ for $k = 1,...,K/2$.}
{With the pair formed, carefully beamforming vectors selection is required. Hence, in NOMA we assume that the {\it beamforming vector} for paired users is the same, {\it i.e.,} $\mathbf{g}_{k} = \mathbf{g}_{K+1-k}$ for all $k = 1,...,K/2$.}

\vspace{2mm}
\noindent\textit{{Assumption 2:}} {In \textit{user-pairing} procedure, we assume that the paired users are aligned with the BS so that the same beamforming can serve all paired users simultaneously.} Hence, by admitting that each pair of devices is spatially aligned with the BS, and using localizing tools described, for instance, in \cite{Zafari2019,Mohamed2020}, one should assume further {\it a priori} \textit{user-pairing} step in NOMA systems.

\vspace{4mm}
\noindent \textit{{Assumption 3:}} In NOMA system, beamforming serves more than one aligned device simultaneously; specifically, in this paper, two aligned devices per cluster are admitted according to the \textit{user-pairing} step, while eliminating the inter-cluster interference (favorable propagation) under adopted perfect \ac{CSI} conditions. 

In this work we adopt the linear \ac{ZF} precoding as defined by the vector:
\begin{align}
\label{eq:zero-forcing}
\mathbf{g}_{k} = \mathbf{h'}_{k}(\mathbf{h'}_{k}^{\text{T}}\mathbf{h'}_{k})^{-1},
\end{align}
and satisfying $\mathbf{h'}_{i}^{\text{T}} \mathbf{g}_{k} = 0, \, \forall \,\, i \neq k$, {\it i.e.}, the favorable propagation effect between users belonging to distinct clusters.

\section{SE-EE in NOMA and mMIMO systems}\label{sec:EE-SE} 
We discuss the SE and EE configurations in the NOMA and mMIMO systems. The operation of the NOMA system requires pairing devices so that the channel coefficients of the devices in the same cluster must be appropriately different, enabling power domain usage. As already mentioned, the interference cancellation process via beamforming presents problems that we will demonstrate below.

\subsection{Data Rates in NOMA with ZF} \label{sec:Data_NOMAZF}
\begin{figure}[!htbp]
\centering
\includegraphics[trim=.3cm 7cm .3cm 7cm, clip, width=.9\linewidth]{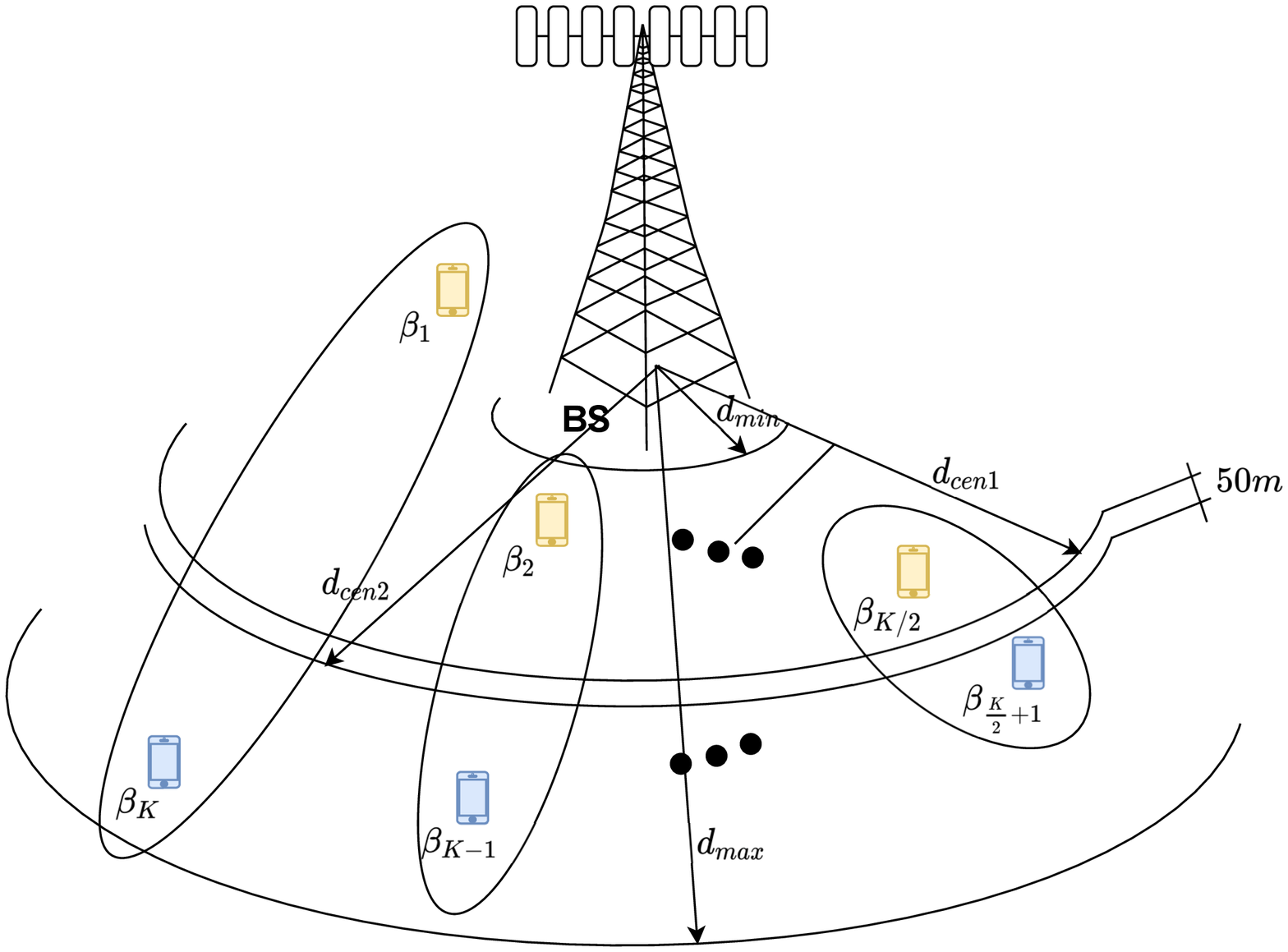}
\caption{System Model indicating the Paring formation in NOMA system. {Both mMIMO and NOMA systems deploy the same massive number of antennas at base-station, $M$.}}
\label{fig:parformation}
\end{figure}

Devices are divided {into} two sets like described in Eq. (\ref{eq:sets}), and these groups are represented by Eq. (\ref{eq:B_order}) by their large-scale fading coefficient and are grouped into pairs forming a cluster as Fig. \ref{fig:parformation}, the cluster $k$ is formed by one device in cell center set $\mathcal{K}_H$ and one device in cell edge set $\mathcal{K}_L$. Hence, the devices are grouped {into} two subsets:
\begin{align} \label{eq:B_order}
\mathcal{K}_H = & \{\beta_1 > \beta_2 > ... > \beta_{k/2}\}, \quad \text{(center devices set)}\\
\mathcal{K}_L = & \{\beta_K < \beta_{k-1} < ... < \beta_{k/2 + 1}\}. \,\,\, \text{(edge devices set)}\nonumber
\end{align}

The {\textit{user-pairing}} adopted in Eq. \eqref{eq:B_order} is {the same} as proposed in \cite{Ali2017}, creating the largest possible difference in channel coefficients for devices not yet paired. 

\vspace{2mm}
\noindent\textit{{Assumption 4:}} {In this paper, we assume perfect} \ac{SIC}, and only one perfect \ac{SIC} stage per cluster is performed, since just 2 devices per cluster are admitted.

\vspace{1mm}
The instantaneous \ac{SINR} of devices in cluster $k$ is defined as:

\begin{equation} \label{eq:SINR}
\text{SINR}_{k}=\frac{\beta_{k}p_{k}|\mathbf{h'}_{k}^{\text{T}} \mathbf{g}_{k}|^{2}}{\beta_{k} \sum^{K}_{k' \neq k}p_{k'}|\mathbf{h'}_{k}^{\text{T}} \mathbf{g}_{k'}|^{2}+1}.
\end{equation}

{In each cluster, the cell-edge devices treat the interference as noise and decode their} data symbols, whereas the cell-center device can decode the data symbols of the cell-edge device and perform \ac{SIC}, hence effectively removing the interference due to the cell-edge device under \textit{{Assumption 3}.}  

{To perform \ac{SIC}}, the cell-center device needs to be able to decode data signal intended for the cell-edge device, {\it i.e.}, the ergodic \ac{SINR} of the cell-edge device, {$\textsc{sinr}_{K+1-k}$, at device $k$, defined as $\textsc{sinr}_{k,K+1-k}$,} must be greater than or equal to the ergodic \ac{SINR} of the $k$-th cell-center device. Hence, {given} the uplink (mMIMO and NOMA) and downlink (NOMA) pilot overhead and assuming perfect CSI in all receivers, and admitting  \textit{{Assumption 4}}, the following condition must be satisfied \cite{Senel2019}, \cite{Cheng2018}:
\begin{equation} \label{eq:SIC}
\mathbb{E} [\text{SINR}_{k,K+1-k}]\,\, \geq \,\, \mathbb{E} [\text{SINR}_{k}], 
\end{equation}
where
\begin{equation} \label{eq:SINR_SIC}
\text{SINR}_{k,K+1-k} =\frac{\beta_{k}p_{K+1-k}|\mathbf{h'}_{k}^{\text{T}} \mathbf{g}_{k}|^{2}}{\beta_{k} \sum^{K}_{k' \neq K+1-k}p_{k'}|\mathbf{h'}_{k}^{\text{T}} \mathbf{g}_{k'}|^{2}+1}.
\end{equation}
Herein, the condition in \eqref{eq:SIC} must be satisfied by selecting the transmit powers appropriately. 

The achievable ergodic rate of devices in cluster $k$, {\it i.e.} device $k$ in $\mathcal{K}_{H}$ subset and device {$K+1-k$} in $\mathcal{K}_{L}$ subset, under Assumptions {1--4}, is given by the ergodic rate contribution of {user-center} device:
\begin{equation} \label{eq:Rnoma1}
\text{R}^{\textsc{noma}}_{k}=\tau \mathbb{E} \left [ \text{log}_{2} \left (1 + \text{SINR}_{k}\right ) \right],\qquad \forall k \in\mathcal{K}_{H}
\end{equation}
{in ${\rm [bits/s/Hz]}$, and for the user-edge device}:
\begin{equation} \label{eq:Rnoma2}
{\text{R}^{\textsc{noma}}_{K+1-k}= \tau \mathbb{E} \left [\text{log}_{2} \left ( 1 + \text{SINR}_{K+1-k} \right ) \right], \qquad \forall k \in \mathcal{K}_{L}}
\end{equation}
where $\tau = (1-\frac{K \cdot {T_{\textsc{s}}}}{\text{T}})$, is the portion of each channel coherence interval ($\text{T})$ that is used for data transmission.

Assuming perfect channel state information, \ac{ZF} precoding for inter-clusters interference elimination, and using random matrix theory results \cite{Marzetta2016}, the $k$-th cluster \ac{NOMA} achievable rate is obtained plugging eq. \eqref{eq:SINR}, (\ref{eq:Rnoma1}) and (\ref{eq:Rnoma2}):
\begin{align}\label{eq:Ratenoma}
\text{R}^{\textsc{noma}}_{\text{cl-}k}= \tau\mathbb{E}\left[\text{log}_2 \left (1+ \bar{M} \beta_{k} p_{k} \right )\right] + \\
\tau\mathbb{E}\left[\text{log}_2 \left (1 + \frac{\beta_{K+1-k}  p_{K+1-k}}{\beta_{K+1-k} p_k +1} \right), \right] \nonumber%
\end{align}
$\forall k \in \mathcal{K}$ and $\bar{M} = M+1-K/2$.  Hence, the \ac{NOMA} system can operate until $K < 2M-1$. A detailed derivation of the expressions on this section can be found in \cite{Senel2019} and \cite{Cheng2018}.

\subsection{Data Rates in mMIMO with ZF}\label{sec:Data_mimozf}
In the \ac{mMIMO} system with \ac{ZF} precoding the ergodic achievable rate for device $k$ is given by:
\begin{equation} \label{eq:Rmimo}
\text{R}^{\textsc{m-mimo}}_{k}=\tau \mathbb{E}  \left [ \text{log}_{2} \left ( 1 + \text{SINR}_{k} \right ) \right ],\quad {\rm [bits/s/Hz]}
\end{equation}
where $\text{SINR}_{k}$ is defined in (\ref{eq:SINR}). Hence, the above \ac{mMIMO} achievable rate equation becomes:
\begin{equation}\label{eq:rate_mMIMO}
\text{R}^{\textsc{m-mimo}}_{k} = \tau \mathbb{E}  \left [ \text{log}_{2} \left ( 1 + (M-K)  p_{k} \beta_{k} \right)\right], \quad {\rm [bits/s/Hz]},
\end{equation}
{where $(M-K)$ is obtained using random matrix theory, representing the coherent array gain of the received signal \cite{Marzetta2016}}. 
Under linear precoding and combiners, the \ac{mMIMO} system operates consistently when $K < M$. Finally, the {\it average system sum-rate} (avg-sum-rate) is defined simply by:
$$
\mathcal{R}^{\textsc{m-mimo}} = \sum_{k=1}^{K}\text{R}^{\textsc{m-mimo}}_{k} \qquad \text{and} \qquad \mathcal{R}^{\textsc{noma}} = \sum_{k=1}^{\mathcal{K}_H} \text{R}^{\textsc{noma}}_{\text{cl-}k}.
$$
The mMIMO system equations have been
thoroughly investigated in literature and can be found in \cite{Marzetta2016} and \cite{Yang2013}.

\subsection{Energy Efficiency}\label{sec: EE-power-terms}
{\ac{EE} metric is the ratio of the number of effective bits of information received over the total energy consumed by the overall system to transmit and receive/decode such information. The system data rate can determine the number of effective information bits received at the destination. Power consumption required for processing the signal at the transmitter and receiver side is often neglected; in this sense, it is calculated just as proportional to the radiated transmitted power. The growth in the number of antennas in the \ac{BS} and the increased number of devices in 5G systems can lead to unattainable \ac{EE} goals. In general, the average \ac{EE} can be expressed as:}
\begin{equation}
\quad \text{EE} =  \frac{\sum^{K}_{k=1}R_k}{P_{\textsc{tot}}}, \qquad {\rm [bits/Joule/Hz]}, \label{eq:EE}
\end{equation}
where $P_{\textsc{tot}}$ is the total power consumption across the communication system. It should consider transmission power consumption, such as RF power amplifier inefficiency, baseband signal processing, and cooling, among others. Therefore, a more realistic and detailed energy consumption model is required.

Based on \cite{Bjornson2014a}, the adopted power consumption model in our work considers two power terms: {\it a}) fixed-term; {\it b}) terms scaled with the number of antennas $M$ and the number of devices $K$. The scaled terms occur because of the transceiver chains, coding/decoding, channel estimation, and precoding. Let the computational efficiency be $L$ operations per joule in BS. We describe it as follows:

\vspace{2mm}
\noindent\textbf{\textit{RF Power}}: $P_{\textsc{rf}}$ is the power consumed {to transmit the signal} to active devices achieved the SINR target and  $0 < \varpi \leq 1$ is the efficiency of the power amplifier.

\vspace{2mm}
\noindent\textbf{\textit{Fixed consumption}}: $P_{\textsc{fixed}}$ is the power consumed at the BS which is independent of the number of transmit antennas and devices in the cell, is formed of term $P_0$ including the power consumption of backhaul {infrastructure}, control signaling, baseband processor, and term $P_{\textsc{syn}}$ a single oscillator used in all BS. 
$$
P_{\textsc{fixed}}=P_0+P_{\textsc{syn}}
$$

\vspace{2mm}
\noindent\textbf{\textit{Dependence only on K}}: $P_{\text{K}}$ is formed by the consumption to coding and modulation of information symbols to devices, { represented by $P_{\textsc{cod}}$}, the consumption to BS decoded the $K$ sequences of information symbols, {defined by $P_{\textsc{dec}}$, and the received power, represented by $P_{\text{RX}}$, still composes this term, multiplied by $K$ as well. In addition, a portion of the ZF precoding cost \cite{Boyd} depends only on $K^3$.} 
$$
P_{\text{K}} = K (P_{\textsc{cod}} +P_{\textsc{dec}}+P_{\text{RX}}) + K^3 \frac{2}{3LT}
$$

\vspace{2mm}
\noindent\textbf{\textit{Dependence only on M}}.  {The term $P_{\text{M}}$ represents the transmitted power} ($P_{\text{TX}}$), hence 
$$
P_{\text{M}} = M P_{\text{TX}}
$$

\vspace{2mm}
\noindent\textbf{\textit{Dependence on K and M}}:  the term $P_{\text{KM}}$ {is the cost of the ZF precoding (due to LU-based matrix inversion) \cite{Boyd}, which depends on the number of devices, the number of antennas, and the vector information symbol.} $$
P_{\text{KM}} = MK \frac{3 + T}{TL} + MK^2 \frac{2}{TL}
$$

{Adding the portions, we obtain the overall power consumption of the system:}
\begin{equation}\label{eq:powermodel}
P_{\textsc{tot}} = \frac{P_{\textsc{rf}}}{\varpi} + P_{\textsc{fixed}} + P_{\text{K}} + P_{\text{M}} + P_{\text{KM}} \quad [\rm W].
\end{equation}


\subsection{Power Allocation Strategies}\label{sec:PAstrateg}

In the sequel, we present three well-known and frequently applied strategies for {power allocation}. Still, due to the inherent characteristics of \ac{NOMA}, we propose modifications on the classical water-filling (WF) algorithm to enable application in the NOMA system.  Such modifications, namely $\Delta$-WF, {ensure that none of the paired devices are dropped-out without undoing the pairing of devices.} To guarantee a certain level of power disparities in {each paired device}, the power allocation $\Delta$-WF procedure in {the \ac{NOMA} system} has two steps: a) first, we allocated power for the clusters; b) we allocate power between paired devices. Thereby, we could analyze the behavior of the systems and compare their results. 

{Notice that both mMIMO and NOMA systems deploy the same massive number of antennas at base-station, $M$. Hence, due to the {\it channel hardening} effect \cite{Marzetta2016, Larsson2014} inherent to massive MIMO configuration, the small-scale fading vanishes across the $M$ antennas equipped with a linear ZF precoding with vector as eq. (\ref{eq:zero-forcing}). Hence, one can consider just the {\it pathloss coefficients} $\beta_k$ as the main parameter in the power allocation policies of systems based on a massive number of antennas.}

\vspace{1mm}
\subsubsection{Equal Power Allocation (EPA)} \label{SEC_EPA}

\ac{EPA} power allocation is deployed as a simple, naive strategy, where all devices are served with {the same power.} In \ac{mMIMO}, all devices are served with the same transmission power regardless of their distance from the \ac{BS}. In \ac{NOMA},  power allocation has two steps. In the first step, the power is allocated equally between the pairs, and then we allocate each device's power equally. The \ac{EPA} strategy applied to \ac{mMIMO} can be defined by:
\begin{equation} \label{eq:EPA}
p_{k} = \frac{P_{\textsc{rf}}}{K} \quad [\rm W], \quad \forall \, k \in \mathcal{K}.
\end{equation}
In the case of \ac{EPA} procedure applied to \ac{NOMA}, it is composed {of two steps}: in the first step, power reference to each cluster can be defined simply as:
\begin{equation} \label{eq:EPA_noma}
p^{\text{cl}}_{\rm ref} = \frac{2\cdot P_{\textsc{rf}}}{K} \quad [{\rm W}], \quad \forall \, k \in \mathcal{K}_H .
\end{equation}
In the second step, the power allocation among the devices in the same cluster is defined as:
\begin{align} \label{eq:EPA_noma2}
p_{k}^{\text{cl-}k} = p_{K+1-k}^{\text{cl-}k} = \frac{p^{\text{cl}}_{\rm ref}}{2} .
\end{align}

\vspace{1mm}
\subsubsection{Proportional Channel Inversion Power Allocation (PICPA)}  \label{SEC_PICPA}
is another power allocation technique adopted in this study. Unlike the EPA technique, which applies the same power to all devices, PICPA applies more power to devices with the worst channel conditions, favoring fairness across the devices{. Such power} allocation penalizes the average sum rate in favor of {fairness among all users.} 

The \ac{PICPA} strategy applied to mMIMO can be defined as:
\begin{equation} \label{eq:PICPA}
p_{k} = P_{\textsc{rf}} \frac{\beta^{-1}_k}{\sum_{k=1}^{K}\beta^{-1}_k} \quad [{\rm W}], \quad \forall k \in \mathcal{K},
\end{equation}
while the \ac{PICPA} procedure applied to \ac{NOMA} follows two steps; in the first step, the power is allocated equally among the $K/2$ clusters:
\begin{equation} \label{eq:PICPA_noma}
p^{\text{cl}}_{\text{ref}} = \frac{2\cdot P_{\textsc{rf}}}{K} \quad [{\rm W}], \quad \forall k \in \mathcal{K}_H,
\end{equation}
after that, the power of each device within the $k$-th cluster is defined by allocating more power to the device with smaller large-scale fading $\beta_k$:
\begin{equation} \label{eq:PICPA_noma2}
p_{k}^{\text{cl-}k} = \,\, p^{\text{cl}}_{\text{ref}} \frac{\beta_{K+1-k}}{\beta_k - \beta_{K+1-k}}, \quad \text{and} \quad
p_{K+1-k}^{\text{cl-}k} = p^{\text{cl}}_{\text{ref}} -p_{k}^{\text{cl-}k},   
\end{equation}
where 
$p_{k}^{\text{cl-}k}$ is the power allocated to the device $k$ in the $\text{cl-}k$ cluster, and $p_{K+1-k}^{\text{cl-}k}$ is the power allocated to the device $(K+1-k)$, also belonging to the $k$-th cluster.

\vspace{2mm}

\subsubsection{Classical Water-Filling (WF) Algorithm} \label{SEC_cWF}
The application of the \ac{WF} algorithm in \ac{mMIMO} system results in {an optimal} (maximum) system sum-rate solution. However, some devices are dropped out of the service due to the deteriorated channel condition. The WF power allocation strategy for \ac{mMIMO} is described as:
\begin{align} \label{eq:WF_mMIMO}
\mu =& \frac{1}{|\mathcal{K}|} \left ( P_{\textsc{rf}} + \sum^{|\mathcal{K}|}_{k=1, k\in\mathcal{K}} \frac{1}{\beta_k} \right ), \\ \nonumber
P_{\textsc{rf}} = \sum^{|\mathcal{K}|}_{k=1, k\in\mathcal{K}} p_k\,, \quad &\text{where} \quad p_k = \left ( \mu - \frac{1}{\beta_k} \right )^+, \forall k \in \mathcal{K} \\ \nonumber 
\text{and} &\quad  {\bf p} = [p_1,p_2,...,p_{|\mathcal{K}|}],
\nonumber
\end{align}
with the operator $(z)^+ = \max(0,z)$. Notice that the constrained value for the total power available is set to $P_{\textsc{rf}} \,\, \rm[W]$. The Algorithm \ref{algo:WF} describes the classical WF procedure.

\begin{algorithm}
\SetAlgoLined
\caption{Classical Water Filling (WF) for mMIMO}\label{algo:WF}
\KwIn{$\mathcal{K}$,P$_{\textsc{rf}}$, $K = |\mathcal{K}|$}
NP $\neq \oslash$\;
\While {(NP $\neq \oslash)$}{solve Eq. (\ref{eq:WF_mMIMO}) 
$\rightarrow$ {\bf p}\;
NP $\gets$ identify null positions in {\bf p}\;
$\mathcal{K} /\{k\}_{\rm NP}$  $\gets$ exclude from {\bf p} devices labeled as NP}
\KwOut{${\bf p} = [p_1,p_2,\ldots,p_{|\mathcal{K}|}]$}
\end{algorithm}

On the other hand, the direct application of \ac{WF} algorithm in the \ac{NOMA} system implies {harming the pair formation}, {\it i.e.,} devices present in {the} $\mathcal{K}_{L}$ set are effectively dropped-out of the service, undoing the pair. We propose a modification in classical WF like the following to allow some comparison.

\vspace{2mm}
\subsubsection{$\Delta$-WF for NOMA} \label{SEC_WFmod}
 
 The application of classical \ac{WF} in NOMA in the same way as it is applied to mMIMO causes some formed pairings to be broken, since after WF algorithm application, some devices are dropped-out from the system, making the NOMA power difference ($\Delta$) in the devices of the same cluster vanish. Hence, we suggest modifying the classical WF procedure to be applied to NOMA accordingly. The steps of the $\Delta$-WF algorithm are described as follows. 
 
 {In NOMA, the power allocation has two steps, in the first step, the allocation is between clusters.} Hence, to prevent the formed pairs from being broken, we propose the application of \ac{WF} based on the large-scale fading differences {of the paired devices, as defined in eq. \eqref{eq:DeltaBeta}:} $\Delta \beta_k= (\beta_k- \beta_{K+1-k})$ inside each $\mathcal{K}_L$ and $\mathcal{K}_H$ subsets, eq.  \eqref{eq:B_order}. In the second step of the procedure, the power is allocated between the devices inside the group, assuming perfect successive interference cancellation (SIC); for this to be possible, the condition in Eq. \eqref{eq:SIC} must be satisfied. {The new water-level in the modified $\Delta$-WF power allocation strategy for NOMA is defined by}
\begin{align} \label{eq:WF_NOMA_new1}
\mu =&  \frac{2}{\mathcal{K}_H}  \left ( P_{\textsc{rf}} + \sum^{|\mathcal{K}_H|}_{k=1,k\in\mathcal{K}_H} \frac{1}{{\Delta\beta_k}} \right ), \\
& P_{\textsc{rf}} = \sum^{ |\mathcal{K}_H|}_{k=1,k\in\mathcal{K}_H}p_{\text{cl-}k}, \qquad \forall k \in \mathcal{K}_H  \notag\\ &\text{where} \quad p_{\text{cl-}k} = \left ( \mu - \frac{1}{{\Delta\beta_k}} \right )^+, \nonumber  \\ 
& \text{and} \quad  {\bf p}_{\text{cl-}k} = [p_1,p_2,...,p_{|\mathcal{K}_H|}], \nonumber
\end{align}
In the second step, the power allocation to both devices in the $k$-th cluster is defined as:
\begin{align} \label{eq:WF_noma2}
p_{K+1-k}^{\text{cl-}k}  = p_{k}^{\text{cl-}k} = {\frac{p_{\text{cl-}k}}{2}}
\end{align}
{Algorithm \ref{algo:D-WF} summarize the proposed $\Delta$-WF power allocation procedure, aiming to improve the SE of NOMA systems.}

\begin{algorithm}
\SetAlgoLined
\caption{$\Delta$-WF (modified) for NOMA systems} \label{algo:D-WF}
\KwIn{$\mathcal{K}_H$, $\mathcal{K}_L$, P$_{\textsc{rf}}$}
NP $\neq \oslash$\;
\While {(NP $\neq \oslash)$}{solve Eq. (\ref{eq:WF_NOMA_new1}) $\rightarrow$ {\bf p}$_{\text{cl-}k}$\;
NP $\gets$ identify null position in {\bf p}$_{\text{cl-}k}$\;
$\mathcal{K}_H /\{k\}_{\rm NP}$ $\gets$ exclude from {\bf p}$_{\text{cl-}k}$ devices labeled as NP
}
\KwOut{{\bf p}$_{\text{cl-}k} = [p_1,p_2,\ldots,p_{|\mathcal{K}_H|}]$}
\end{algorithm}

\vspace{2mm}
\noindent{\noindent{\bf Complexity analysis:} In a comparative analysis of complexity, {the $\Delta$-WF algorithm} for power allocation in NOMA system (Algorithm \ref{algo:D-WF}) performs two simple additional operations compared to the classical WF procedure (Algorithm \ref{algo:WF}):
a) in eq. (\ref{eq:WF_NOMA_new1}) the subtraction in ${(\beta_k - \beta_{K+1-k})}$; and b) the division by two in (\ref{eq:WF_noma2}). Besides, NOMA {\it vs.} mMIMO systems require different {\it a priori} information to proceed accordingly with the PA procedure.}

\subsection{{Prior Information for Power Allocation Step}} \label{sec:Previous2}
{For implementing the \ac{PA} policies, prior information is required at the BS, as defined in Table \ref{tab:prior}.  Some of this necessary information can be obtained through the dedicated {\it pilot} transmission step, at the cost of some overhead, as described in Section \ref{sec:model_pilot}.  Moreover, a preliminary step is known, in which the {\it spatial localization} and {\it path loss} estimation of all devices must be realized. With such {\it a priori} information availability,  the {\it user-sorting} and {\it user-pairing} steps can be performed.}

\begin{table}[!htbp]
\centering
\caption{{Prior information required to PA procedure}}
\begin{tabular}{|cc|c|c|c|c|c|l}
\hline
\multicolumn{2}{c|}{\textbf{PA}}& $\beta_k$& $h_k'$& K &$\mathcal{K}_H$ & $\mathcal{K}_L$ & $p_k$, eq.  \\ \hline \hline
\multicolumn{1}{c|}{\multirow{2}{*}{EPA}}  & mMIMO &$-$& $-$   &$\blacktriangle^*$ &$-$  & $-$  & (\ref{eq:EPA}) \\ \cline{2-8} 
\multicolumn{1}{c|}{}                      & NOMA  &$-$& $-$   &$\blacktriangle$ &$-$  & $-$  & (\ref{eq:EPA_noma}), (\ref{eq:EPA_noma2})     \\ \hline
\multicolumn{1}{c|}{\multirow{2}{*}{PICPA}}& mMIMO &$\blacktriangle^*$& $-$   & $-$ & $\blacktriangle^*$ & $\blacktriangle^*$ & (\ref{eq:PICPA})      \\ \cline{2-8} 
\multicolumn{1}{c|}{}                             & NOMA  &$\blacktriangle$& $-$   & $-$ & $\blacktriangle^*$      &$\blacktriangle$ & (\ref{eq:PICPA_noma}), (\ref{eq:PICPA_noma2})    \\ \hline
\multicolumn{1}{c|}{WF}          & mMIMO &$\blacktriangle^*$& $-$   & $-$ & $\blacktriangle^*$  & $\blacktriangle^*$       & (\ref{eq:WF_mMIMO}), Alg. 1 \\ \hline
\multicolumn{1}{c|}{$\Delta$-WF} & NOMA &$\blacktriangle$& $-$    & $-$ & $\blacktriangle^*$  & $\blacktriangle$       & (\ref{eq:WF_NOMA_new1}, \ref{eq:WF_noma2}), Alg. 2     \\ \hline 
\end{tabular}\\
{\flushleft $\blacktriangle$ information needed a prior \qquad  $^*$ obtained via Pilot Overhead \\}
{\flushleft $-$ Information not needed \\}
 \label{tab:prior}
\end{table}

\section{Numerical Results}\label{sec:results}
The numerical evaluations for the proposed analyses of NOMA {\it and} mMIMO systems are presented in this section. The simulation system and channel parameter values deployed along this section are depicted in Table \ref{tab:simul}. The \ac{BS} is located at the center of cell and equipped with massive $M$ BS transmit antennas in typical non-line-of-sight (\ac{NLOS}) channel propagation scenario. At the same time, the devices are randomly distributed in the cell area and split into two subsets, $\mathcal{K}_L$ and $\mathcal{K}_H$, as illustrated in Fig. \ref{fig:parformation}. In all simulations, we consider a block fading model where the time-frequency resources are divided into coherence time intervals (\text{T}), in which the channels remain constant and frequency flat, and it is measured in multiple of symbol transmit period ($T_\textsc{s}$). The system and channel scenarios have been simulated using Matlab 2019 software running under one Intel HD Graphics 6000 GPU, Intel(R) Dual-Core(TM) I5 CPU @ 1.6 GHz, and 8 GB RAM.

\begin{table}[!htbp]
\centering
\caption{\\Simulation Parameters}
\begin{tabular}{l|l}
\hline
\textbf{Parameter}    & \textbf{Value}           \\ \hline \hline
BS antennas & $M = 64, 128$ and $256$           \\ 
Max. \# Devices in the cell & $K = \zeta\cdot M$ (NOMA)\\
& $K = M$ (mMIMO) \\ 
Cell loading                & $\rho$ = $K/M$ \\
Total RF power available & $P_{\textsc{rf}} = 1 \rm W$\\
Pairs NOMA / Clusters		 & $N = K/\zeta = K/2$ 	\\
NOMA devices per cluster & $\zeta=2$\\
\# antennas per device        	 & 1             			\\ 
Cell edge length   & $d_{\max} = 350$ m                    \\ 
Strong device position & $[d_{\min};\, d_1] \in [50; 100]$ m  \\ 
Weak device position  & $[d_2; \, d_{\max}] \in [150; 350]$ m              \\ 
Array gain MIMO device		 & $M-K$ 		\\ 
Array gain NOMA $k_H$ 		 & $M+1-K/2$ \\ 
Data symbol period  & $T_\textsc{s}$\\ 
Coherence time interval & $\text{T}= 512 \cdot T_\textsc{s}$, \,\, $\iota=512$	\\
\hline
\multicolumn{2}{c}{ \bf Channel}\\
\hline \hline
Pathloss exponent       	 & $\xi = 3.78$   \\ 
Attenuation at a $d_0$ reference & $\beta_0=130$ [dB]\\
{\# Monte-Carlo realizations} 	 & {1000}	\\
\hline
\end{tabular}
 \label{tab:simul}
\end{table}

\subsection{Spectral Efficiency Comparison} \label{sec:SE}
\begin{figure}[!htbp]
\centering
\small
\includegraphics[width=.95\linewidth]{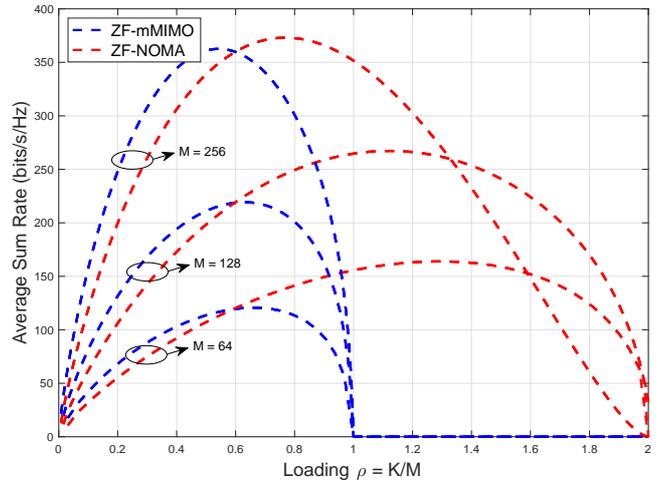} \\
(a) EPA \\
\includegraphics[width=.95\linewidth]{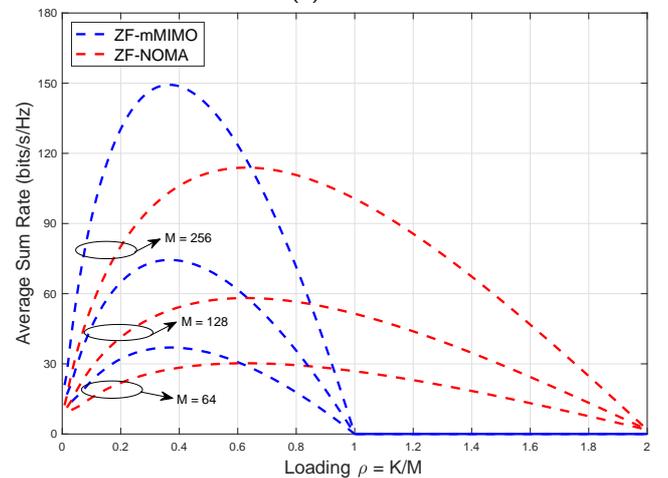}\\
 (b) PICPA\\
\includegraphics[width=.95\linewidth]{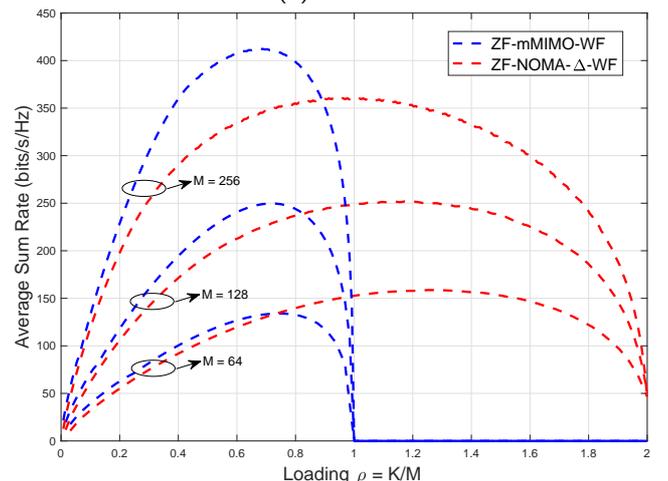}\\
(c) Classical WF in ZF-mMIMO and $\Delta$-WF in ZF-NOMA
\caption{\small The average sum-rate with the loading of devices $0<\rho\leq 2$, considering four power allocation methods: EPA, PICPA, WF, and $\Delta$-WF The average is obtained over 1000 random devices locations.}
\label{fig:SE_ZF_4PA}
\end{figure}

The \ac{mMIMO} and \ac{NOMA} SE performance analysis is carried out in this subsection, by increasing the number of devices two by two until the loading limit $\rho = $ 2. The results consider $M = 64$, $128$ and $256$ BS antennas. In Fig. \ref{fig:SE_ZF_4PA}. (a) the results of SE are achieved when the RF power available is allocated following the EPA strategy, where each device receives the same \ac{PA} values. The mMIMO system overcame NOMA in all situations when the loading $\rho <$ 0.6. However, the NOMA system achieves a higher SE than the mMIMO for each $M$ scenario when {the loading of devices} increases, $\rho>0.6$. The maximum avg-SE is $373$ [bits/s/Hz], being attained with ZF-NOMA $M=256$ antennas and $\rho\approx0.76$. Besides, one can infer that the \ac{mMIMO} does not work with a loading higher than 1, due to the array gain reaching 0 at full {loading of devices}, while NOMA works suitably until the loading attains $M \cdot \zeta$, where $\zeta$ is number of devices per cluster.

Fig. \ref{fig:SE_ZF_4PA}.(b) depicts the results of SE achieved in the mMIMO and NOMA systems when the \ac{PICPA} method is applied to allocate the available RF power per device along the BS antennas. The maximum avg-SE in mMIMO system overcomes the NOMA counterpart until the loading $\rho$ exceeds $\approx 0.62$ for the three BS antenna configurations, $M = 64, 128$, and $256$. This \ac{PA} technique provides more power to devices with the worst channel condition, making the SE result reach maximum values below the EPA.

Fig. \ref{fig:SE_ZF_4PA}.(c) depicts the conventional WF algorithm applied to mMIMO. Under such a power allocation approach, we highlight that forming pairs is unfeasible in the NOMA system. Indeed, the WF algorithm can maximize the system SE since it allocates more power to devices with better channel conditions. In contrast, devices under bad channel conditions (below the water level) are dropped out of the service.

The classical WF algorithm has been adapted to the NOMA system dropped-out always a pair of devices. Such adaptation reveals substantial improvements of avg-sum-rate when $M$ is low compared to classical WF \ac{PA} in mMIMO. The $\Delta$-WF power allocation procedure preserves the pairs clustering formation in the NOMA system, allocating more power to the cluster with a higher difference between coefficients of large-scale fading. Fig. \ref{fig:SE_ZF_4PA}.(c) shows that the maximum avg-SE $\approx 361$ [bits/s/Hz], which is achieved under $\rho = 1$ ($K\approx 256$ devices) when the modified WF is deployed in NOMA system. Moreover, when the number of BS antennas is lower ($M=64$ or $128$), the NOMA achieved a peak higher than mMIMO, e.g., for $M=64$ antennas, the peak of SE mMIMO occurs at loading $\rho\approx 0.7$, while the NOMA SE peaks at $\rho\approx 1.2$. However, as the number of BS antennas grows, the NOMA SE advantage decreases. 

\begin{figure}[!htbp]
\centering
\includegraphics[width=.951\linewidth]{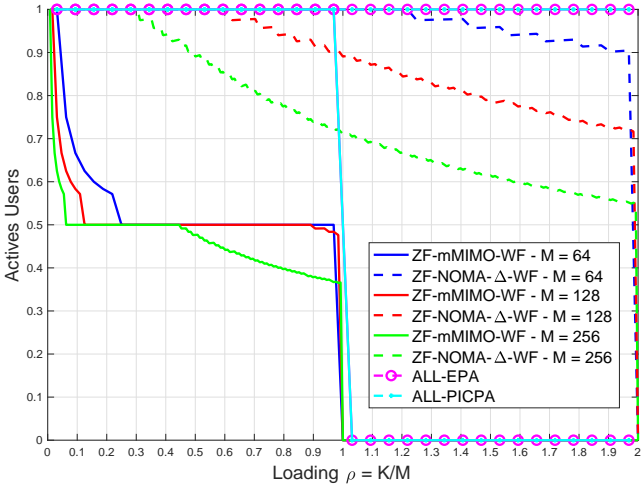}
\caption{The average of active devices after PA procedure {\it versus} {loading of devices} in the range $0<\rho\leq2$. The average is obtained over 1000 random devices locations.}
\label{fig:ActiveUsers}
\end{figure}

\vspace{1mm}
\noindent{\bf Number of active devices after PA procedure}. Fig. \ref{fig:ActiveUsers} shows the number of actives device after applying PA methods: in the EPA and PICPA algorithms, all devices remain activated. However, in classical WF mMIMO system {when the number of device increase}s beyond $\rho \approx 0.25$, half of the devices are dropped-out; while in $\Delta$-WF NOMA \ac{PA}, the percentage of active devices is always higher, e.g. higher than 70\% for $\rho\approx 1.1$ and $M=256$ antennas, {the worst case}.   

\begin{figure}[!htbp]
\centering
\includegraphics[width=.96\linewidth]{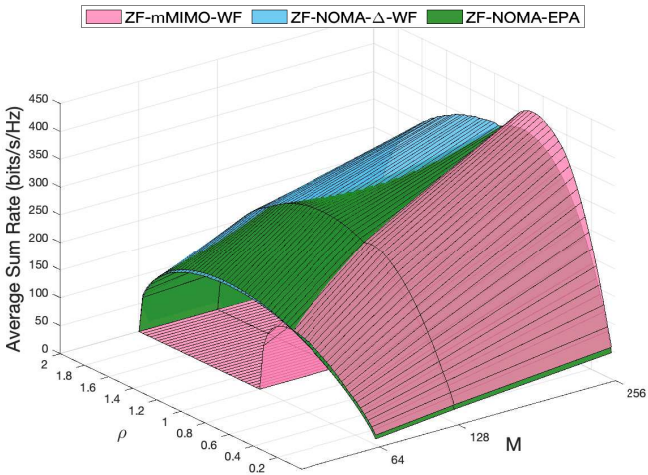}
\caption{Average sum rate with loading $\rho$ and M.}
\label{fig:SE_surf}
\end{figure}

Fig. \ref{fig:SE_surf} summarizes avg-sum-rate surfaces in terms of SE $\times \rho\times M$ results achieved by NOMA with EPA, mMIMO with WF,   and NOMA with modified $\Delta$-WF. In the initial loading part, $\rho<0.65$, the classical WF \ac{PA} in ZF-mMIMO achieves better results until the loading (pink surface). When the number of antennas is low as $M = 64$ and $\rho$ is in between $0.7$ and $1.8$, the EPA \ac{PA}  applied to NOMA (green surface) achieved superior results. Moreover, when $M = 128$ and $0.8<\rho<1.6 $, the ZF-NOMA-EPA achieve superior results (green surface). For a higher number of antennas in BS, {\it i.e.,} $M = 256$ only in a short {loading of devices} range, $ 0.86 <\rho< 0.97$, the ZF-NOMA-EPA achieves superior SE results.  Finally, when $\rho > 0.97$, the modified $\Delta$-WF achieves competitive results (blue surface).

\subsection{Jain's Fairness Index} \label{Jain}
{Another critical analysis developed was to analyze the fairness between the devices, {\it i.e.}, to know the difference in the transmission rate achieved by active devices in the cell. For this measure, we use the Jain's Fairness index like described in \cite{Jacob2020} and can be defined as:}
\begin{equation} \label{eq:Fairness}
\mathcal{F}^{\textsc{syst}}_{\textsc{m}} = \frac{\left(\sum^M_{k=1}R_k \right)^2}{M \sum^M_{k_1}R^2_k}.
\end{equation}

\begin{figure}[!htbp]
\centering
\small
\includegraphics[width=.961\linewidth]{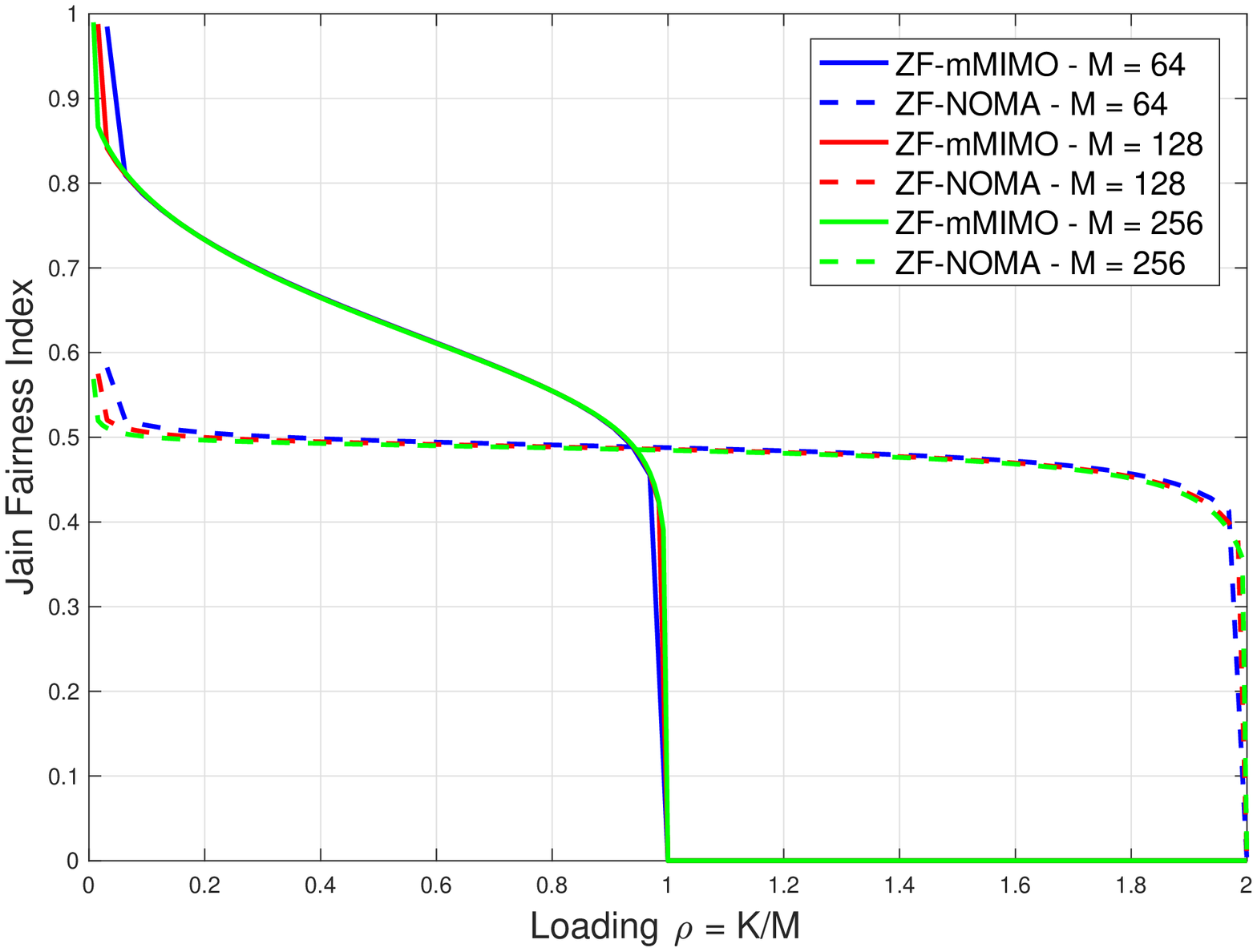}\\
(a) EPA\\
\includegraphics[width=.951\linewidth]{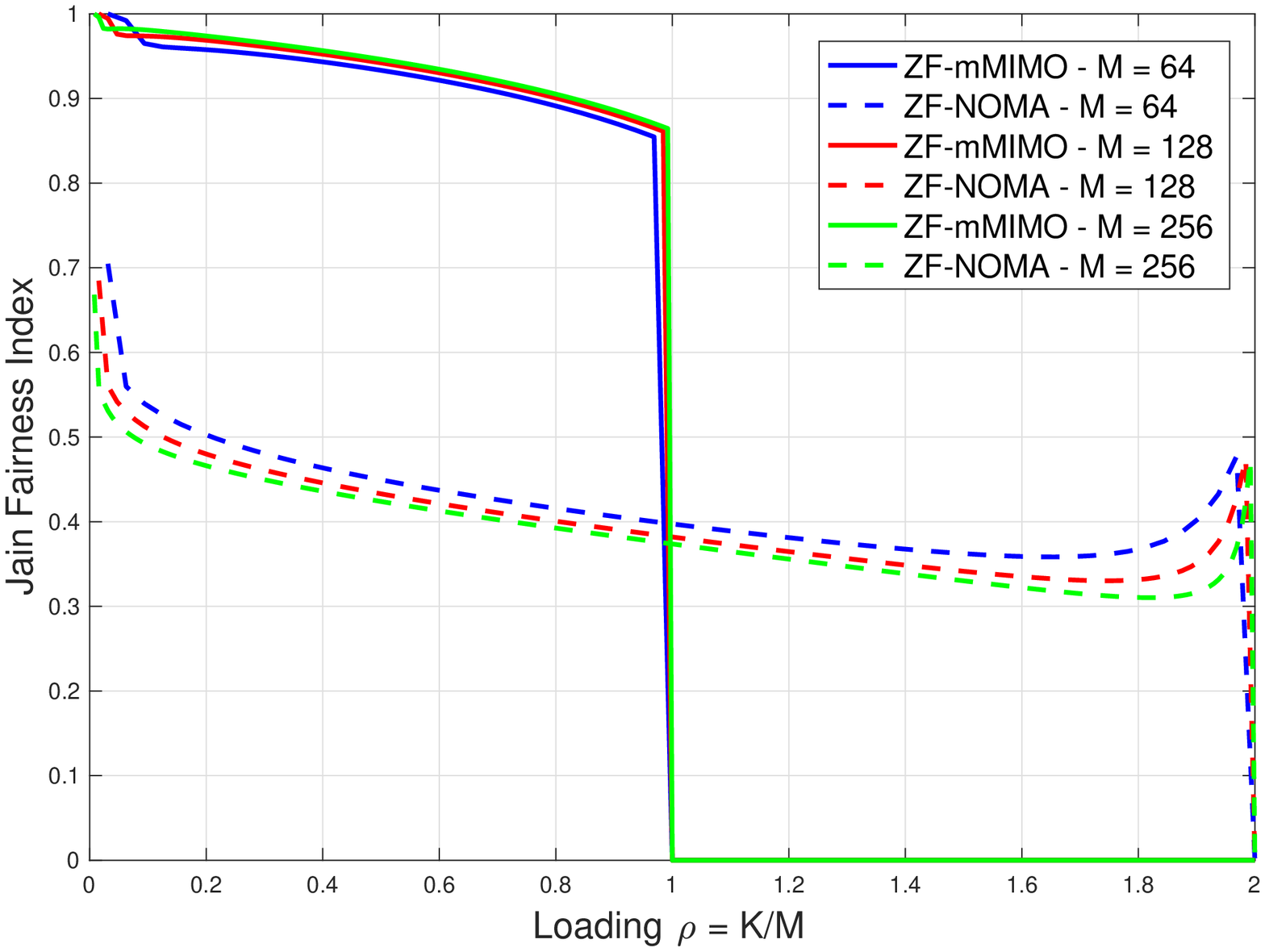}\\
(b) PICPA\\
\includegraphics[width=.9565\linewidth]{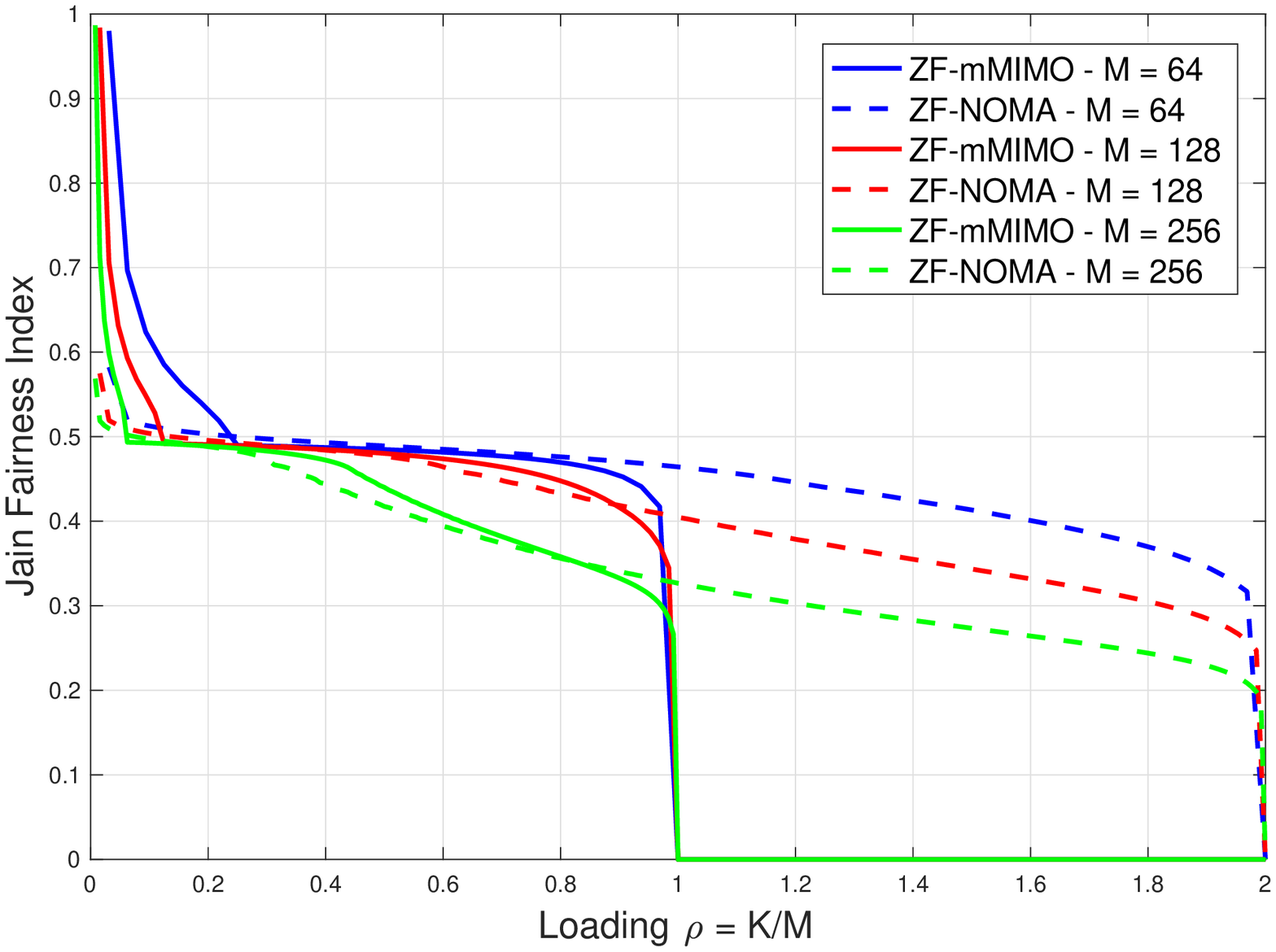}\\
(c) WF and $\Delta$-WF
\caption{\small The fairness of NOMA {\it and} mMIMO system under three power allocation procedures: (a) EPA; b) PICPA; (c) WF and $\Delta$-WF. The average is obtained over 1000 random device locations.}
\label{fig:fairness}
\end{figure}

The Fig. \ref{fig:fairness}. depicts the fairness curves attainable by NOMA and mMIMO systems with EPA, PICPA, WF, and $\Delta$-WF \ac{PA} procedures when the {loading of devices} grows until $\rho = 2$. Fig. \ref{fig:fairness}.(a) shows the Jain's Fairness Index when EPA policy is used, the mMIMO system performs better $\mathcal{F}$ results than NOMA for $\rho < 1$, on the other hand, NOMA can attain $\mathcal{F}^{\textsc{noma}}_\textsc{m} \approx 0.5$ in almost every {loading of devices}, independent of M.

Fig. \ref{fig:fairness}.(b) reveals the Jain's Fairness Index when the PICPA method is applied, despite the SE result being lower in this \ac{PA} method, the mMIMO obtains the best fairness result, keeping the $\mathcal{F}^{\tiny{m}\textsc{mimo}}_\textsc{m}$ consistently above 0.85, still the NOMA under 0.5.

The Jain's Fairness Index when WF and $\delta$-WF are depicted in Fig.\ref{fig:fairness}(c), It is intrinsic to these algorithms to allocate more power to devices with better channel conditions, which causes fairness between devices to be impaired. In this method, it is possible to observe a significant influence of the number of antennas $M$ in the BS and the fairness result.

\subsection{Energy Efficiency Comparison} \label{sec:EE}
Energy efficiency (EE) is another important figure of merit used to compare systems' performance. In this section, a power consumption model based on fixed circuitry power part and that one varying according to the number of antennas $M$ and the number of devices $K$ has been adopted, following eq. \eqref{eq:powermodel}.

\begin{figure}[!htbp]
\centering
\small
\includegraphics[width=.9565\linewidth]{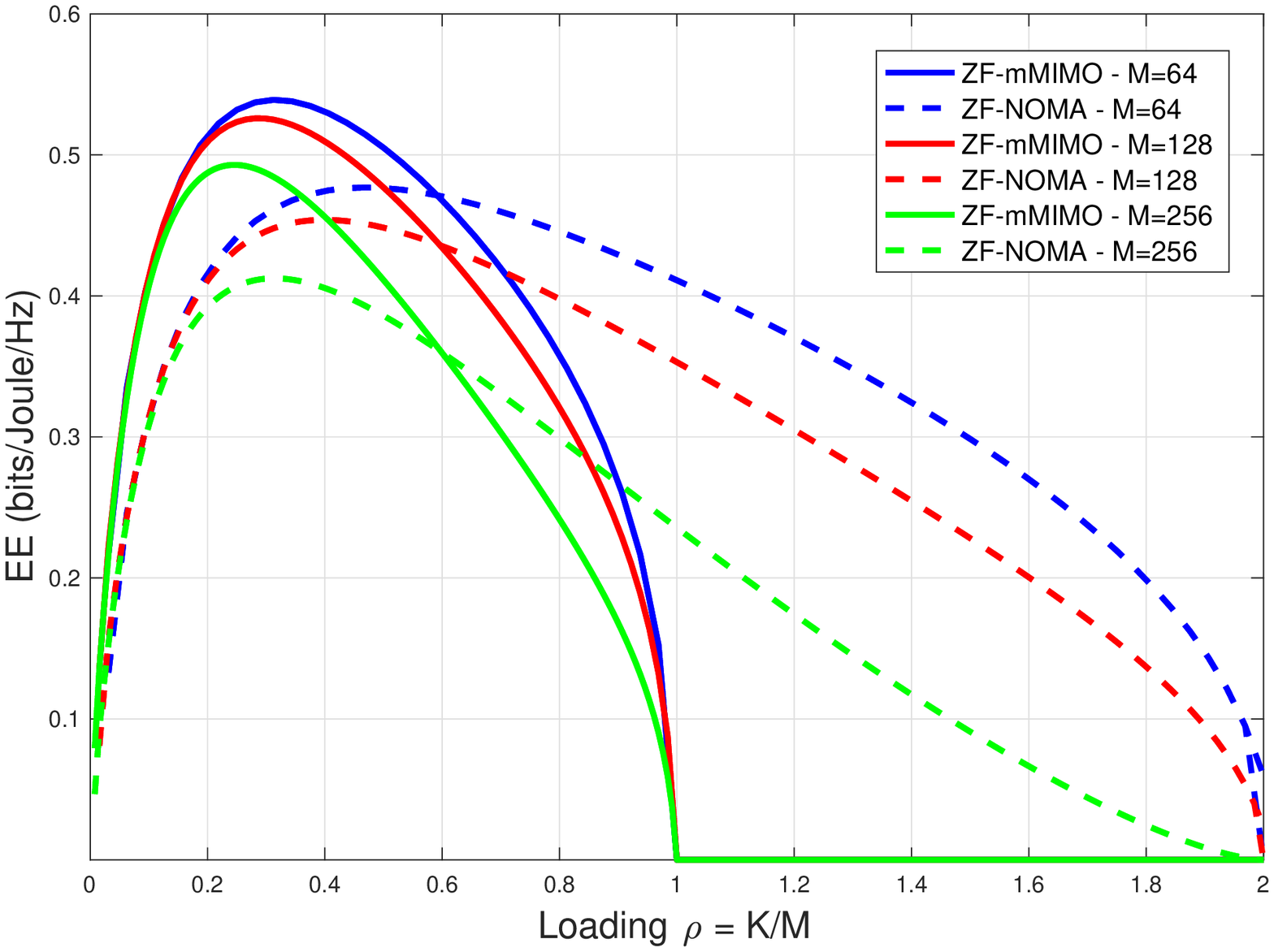}\\
(a) EPA\\
\includegraphics[width=.9565\linewidth]{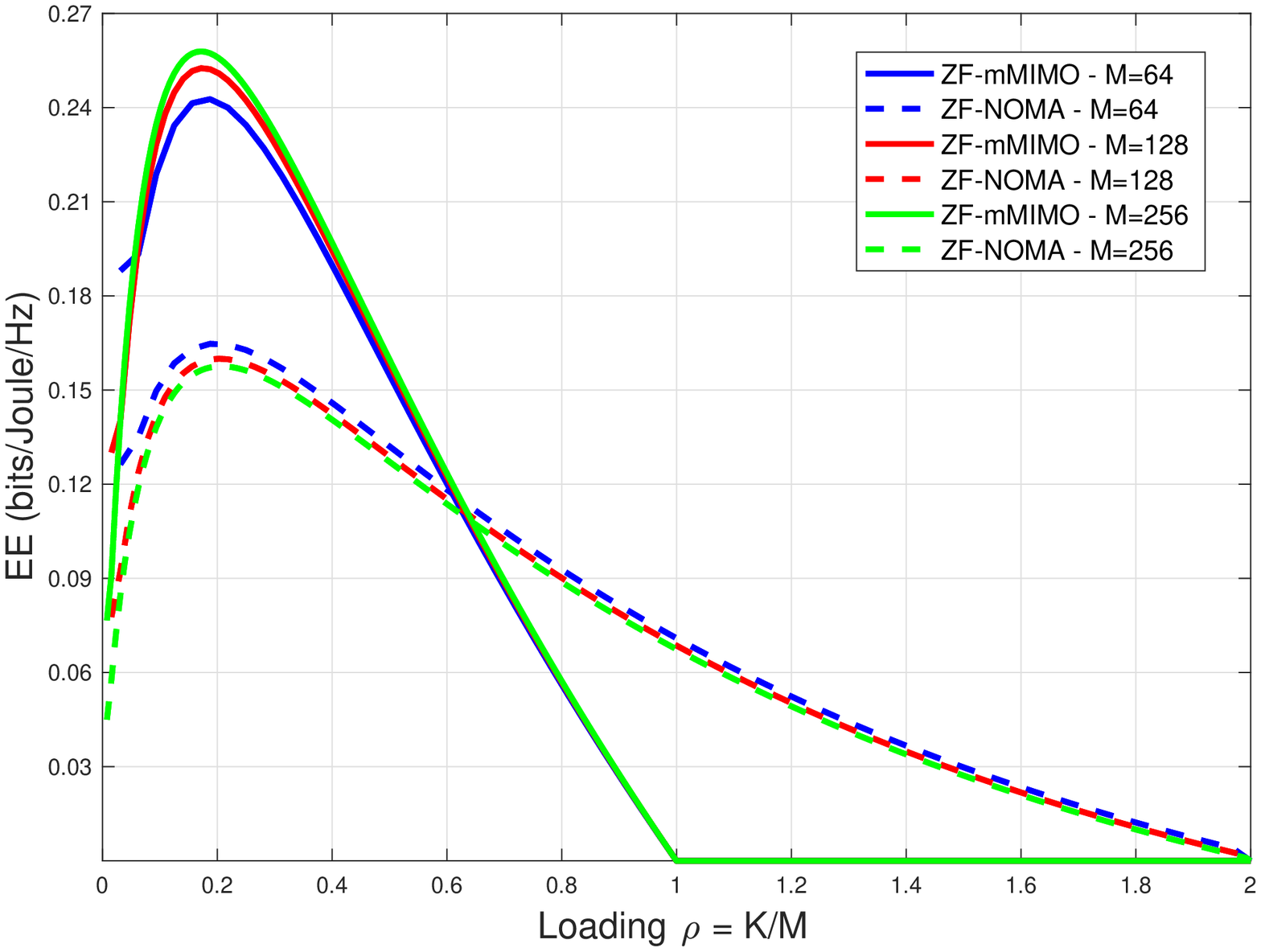}\\
(b) PICPA\\
\includegraphics[width=.9565\linewidth]{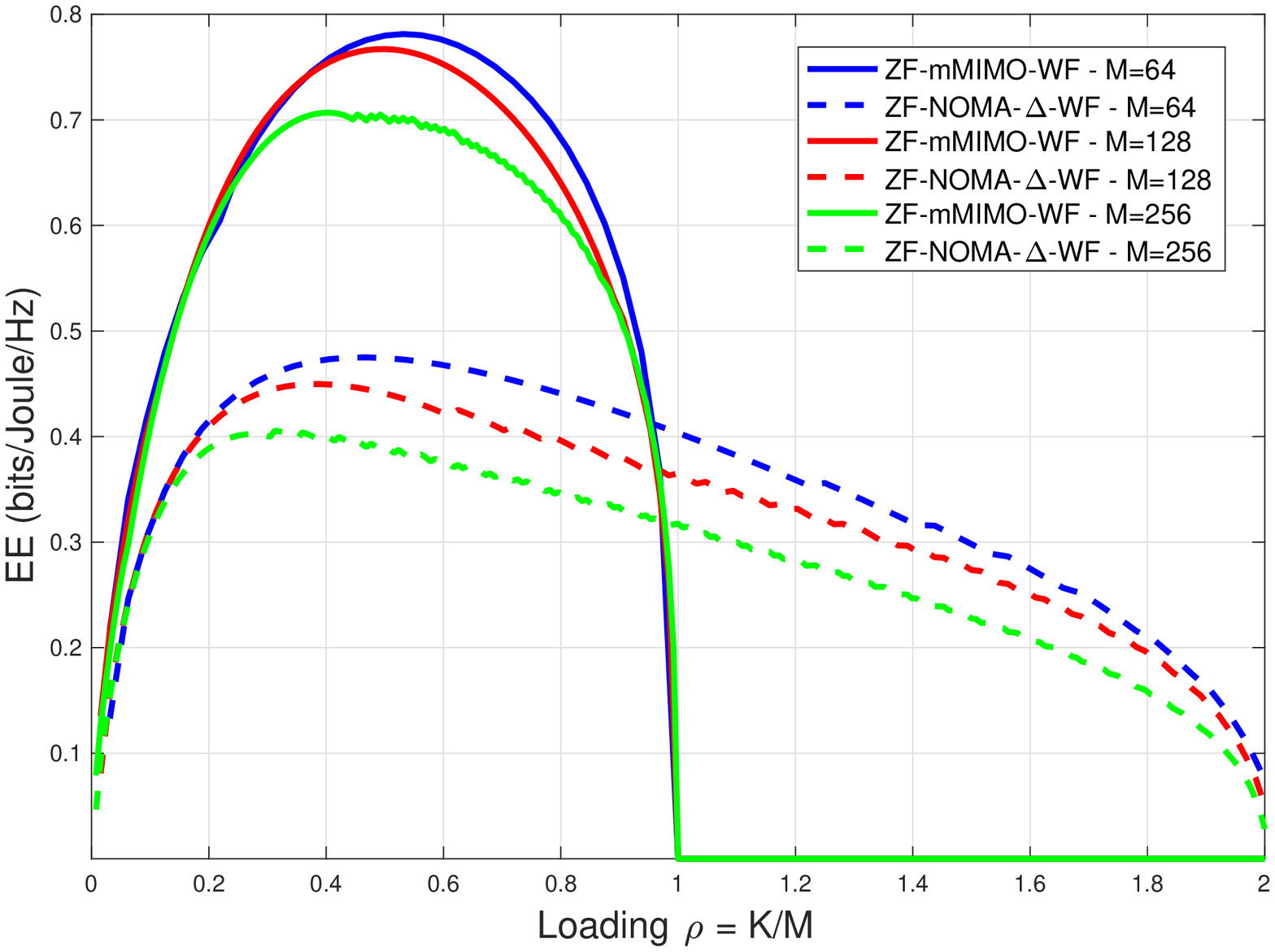}\\
(c) WF and $\Delta$-WF
\caption{\small EE for NOMA {\it vs.} mMIMO  under three power allocation procedures: (a) EPA; b) PICPA; (c) WF and $\Delta$-WF.  Average EE obtained over 1000 random devices locations.}
\label{fig:EE_ZF_4PA}
\end{figure}

Table \ref{tab:EEparameter} {\cite{Bjornson2014a}} presents the adopted parameter values for the EE analysis and comparison discussed in this subsection.  Fig. \ref{fig:EE_ZF_4PA} depicts the performance of EE with EPA, PICPA, WF, and $\Delta$-WF \ac{PA} procedures, considering the exact three quantities of antennas. Fig. \ref{fig:EE_ZF_4PA}.(a) shows the EE performance with EPA. In this method, all devices receive the same power. The avg-EE mMIMO overcomes the NOMA around 13\% to 20\%. It was possible to observe that adding antennas in the BS increases the power consumption, harming the EE result. Again, under {loading of devices} $0.6<\rho \leq 2.0$, the NOMA overcomes the mMIMO system.

\begin{table}[!htbp]
\centering
\caption{Adopted Parameters values for EE analysis \cite{Bjornson2014a}}
\begin{tabular}{l|l}
\hline	\textbf{Parameter}    & \textbf{Value}         \\ \hline \hline
		Backhaul Infrastruture    & $P_0$ = 2 W            \\ \hline
		Single oscillator         & $P_{syn}$ = 2 W  		\\ \hline
		Coding and modulation     & $P_{COD}$ =  4 W per device	\\ \hline
		Decoding and demodulation & $P_{DEC}$ = 0.5 W per device   \\ \hline
		Receive power             & $P_{RX}$ =  0.3 W per device   \\ \hline
		Transmitted power          & $P_{TX}$ = 1 W per antenna     \\ \hline
		Efficiency of Power Amplifier		   & $\varpi=$ 0.3	\\ \hline
		Operations/Joule       	 & L = $10^{9}$ oper. per joule	\\ \hline
	\end{tabular}
	 \label{tab:EEparameter}
\end{table}

Fig. \ref{fig:EE_ZF_4PA}.(b) reveals the EE performance with PICPA. In this method, more power is allocated to devices with poor channel coefficients, resulting in reduced poor EE performance for both NOMA and M-MIMO systems, attaining a maximum of 0.25 and 0.16 [bits/W] for mMIMO and NOMA, respectively. The maximum EE attained by mMIMO is generally around 50\% higher than NOMA. However, for {loading of devices} $\rho\geq 0.65$, NOMA overcomes mMIMO EE performance. 

Fig. \ref{fig:EE_ZF_4PA}.(c) depicts the \ac{EE} performance in the mMIMO with classical WF and in the NOMA with $\Delta$-WF algorithm. It is possible to confirm the superiority of energy efficiency of mMIMO within the range where it operates consistently, {\it i.e.,} $0<\rho < 1$. Notice that the maximum EE achieved by mMIMO is about 70 \% higher than NOMA for different BS antennas. Finally, NOMA becomes more energy efficient than mMIMO only when the {loading of devices} is high, $\rho>0.95$.

In all analyzed system scenarios, the mMIMO equipped with classical WF \ac{PA} procedure achieves higher maximum EE. The mMIMO attains better EE results than NOMA for $\rho<1$. On the other hand, NOMA can serve a more significant number of devices (twice) than mMIMO.

\begin{figure}[!htbp]
\centering
\includegraphics[width=.96\linewidth]{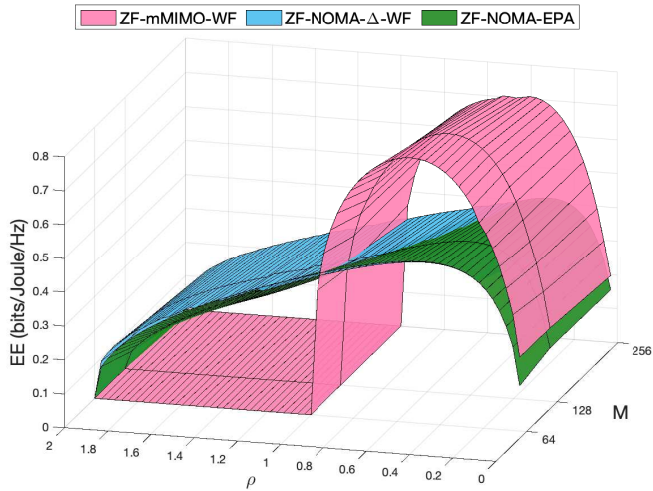}
\caption{EE with loading $\rho$ and M.}
\label{fig:EE_surf}
\end{figure}

Fig. \ref{fig:EE_surf}  summarizes the best EE results in a surface plotting for the mMIMO with WF overcoming NOMA across the entire loading range where it operates consistently. For device loading $\rho > 1$, the NOMA operates under lower EE until the loading $\rho = 2$. Moreover, considering the smallest number of antennas in the BS, the NOMA with EPA overcame the NOMA with modified $\Delta$-WF; despite that, as the number of antennas in the BS increases, the NOMA with $\Delta$-WF achieves marginal superior EE results.

\subsection{\bf Area Under Curves SE and EE}
\begin{figure}[!htbp]
\centering
\small
\includegraphics[width=.9565\linewidth]{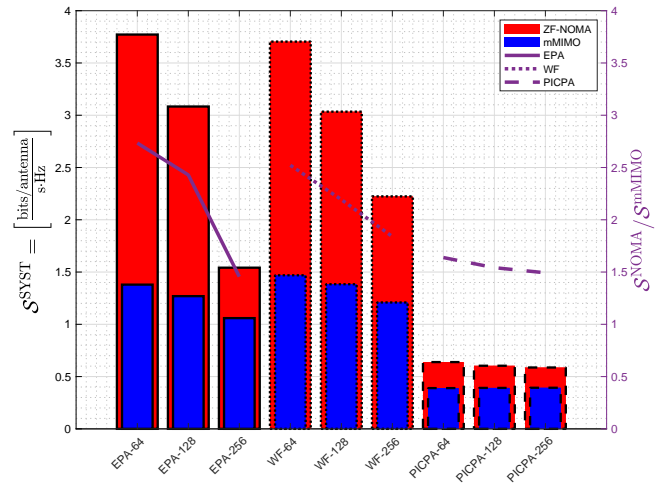}\\
(a) $\mathcal{S}$ - Bar chart of Area Under SE curves. \\
\vspace{2mm}

\includegraphics[width=.9565\linewidth]{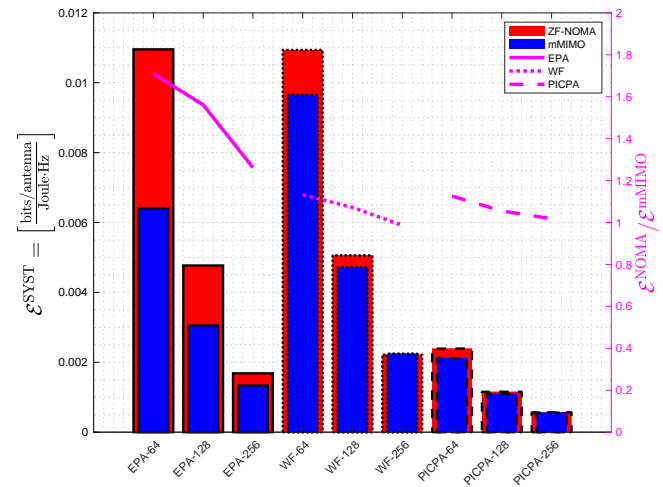}\\
(b) $\mathcal{E}$ - Bar chart of Area Under EE curves.\\
\vspace{2mm}

\caption{The Area Under curve of NOMA {\it and} mMIMO system under three power allocation procedures: (a) SE curves; b) EE curves. The average is obtained over 1000 random devices locations.}
\label{fig:BAR_GENERAL}
\end{figure}

For a fair comparison, one can consider a wide range of average SE and EE along the {loading of devices}, and normalized per antenna, which can be attainable by NOMA and mMIMO systems. Hence, let us consider the corresponding areas under the SE and EE curves in Fig. \ref{fig:SE_ZF_4PA} and Fig. \ref{fig:EE_ZF_4PA}, such that: 
$$
\mathcal{S}^{\textsc{syst}}_{\textsc{M}} = \frac{1}{M}\cdot\int_0^ {\rho=2} \overline{\rm SE}(\rho) \,\, d\rho \qquad \left[\frac{\rm bits/antenna}{\rm s\cdot Hz}\right]
$$ 
and
$$
\mathcal{E}^{\textsc{syst}}_{\textsc{M}} = \frac{1}{M}\cdot\int_0^ {\rho=2} \overline{\rm EE}(\rho)  \,\, d\rho \qquad \left[\frac{\rm bits/antenna}{\rm Joule\cdot Hz}\right],
$$ 
respectively, where $\overline{\rm SE}(\rho) $ is the average overall system sum-rate, and $\overline{\rm EE}(\rho) $ is the average overall system energy efficiency achieved under specific {loading of devices} $\rho$. Hence, comparing the values of corresponding areas under the SE and EE curves of Fig. \ref{fig:SE_ZF_4PA} and Fig. \ref{fig:EE_ZF_4PA}, we obtained  Fig. \ref{fig:BAR_GENERAL}.

From the SE perspective, and considering EPA policy, the higher area-under-SE-curve ratio gain is achieved when the number of BS antennas is $M = 64$:
$$
\mathcal{S}^{\textsc{noma}}_{\textsc{M}=64} \approx 2.7 \cdot  \mathcal{S}^{\tiny{m\text{MIMO}}}_{\textsc{M}=64}.
$$
Notice that when the number of antennas $M$ grows, the ratio above decreases. In the same way, considering WF policy, the gain trend remains. In contrast, considering the PICPA policy, the ratio practically remains the same.

Furthermore, considering now the EE perspective, under EPA policy, the higher ratio is achieved when the BS is equipped with $M=256$ antennas:
$$ \mathcal{E}^{\textsc{noma}}_{\textsc{M}=256} \approx 1.8 \cdot  \mathcal{E}^{\tiny{m\text{MIMO}}}_{\textsc{M}=256}.
$$
As one can conclude, in almost all scenarios, NOMA is more spectrally and energetically efficient than mMIMO over an extensive range of {loading of devices} $0< \rho \leq 2$, roughly, in average, 80\% in terms of energy efficiency, and 170\% more efficient in terms of spectral efficiency.

\subsection{Resource Efficiency (SE-EE Trade-off)} \label{sec:SE-EE}
\begin{figure}[!htbp]
\centering
\includegraphics[width=.99\linewidth]{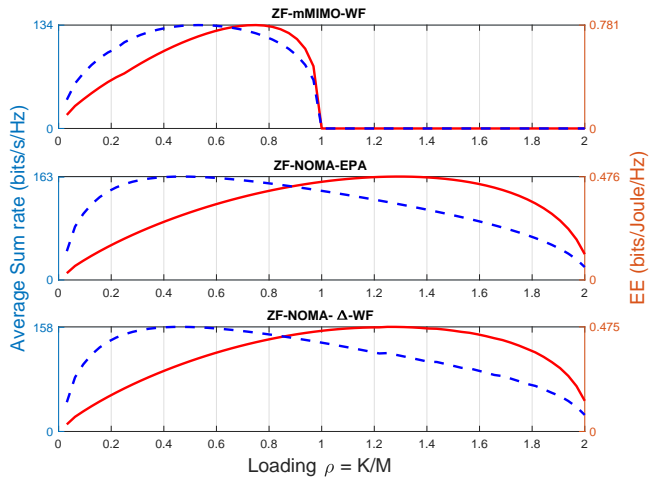}\\
(a) $M = 64$ BS antennas\\
\includegraphics[width=.99\linewidth]{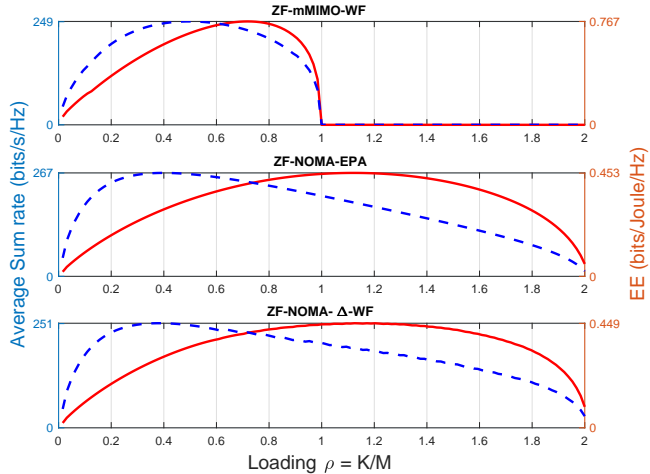}\\
(b) $M = 128$ BS antennas\\
\includegraphics[width=.99\linewidth]{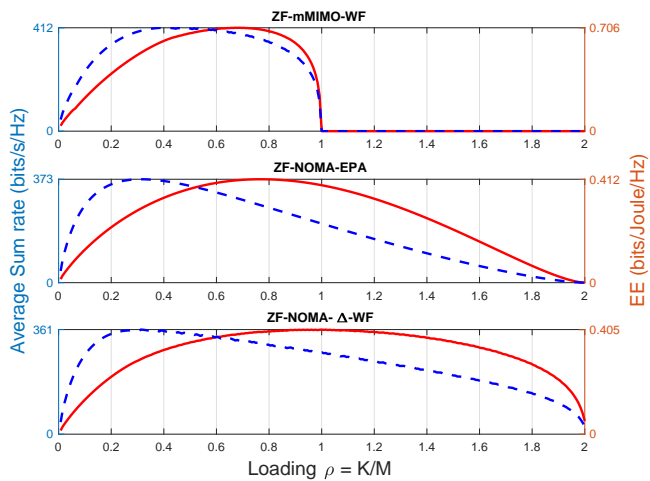}\\
(c) $M = 256$ BS antennas
\caption{SE-EE trade-off points when $M = 64, 128$ and $256$.}
\label{fig:SE-EE-compAll}
\end{figure}

The \ac{NOMA} and \ac{mMIMO} are analyzed in terms of SE and \ac{EE} trade-off, namely {\it resource efficiency} (RE), considering {loading of devices} increasing up to 2. From Fig. \ref{fig:SE-EE-compAll}, one can find graphically the best {loading of devices} range that maximizes the SE-EE trade-off for each BS antenna configuration $M$. The left y-axis depicts avg-SE, and the right y-axis shows the avg-EE. Table \ref{tab:table3} summarizes the optimal {loading of devices} that maximizes the SE-EE trade-off and shows the percentage of active users after Power Allocations and Jain's Fairness Index. Fig. \ref{fig:SE-EE-compAll}.(a) reveals the results when $M = 64$, NOMA with EPA achieved SE-EE trade-off with the highest {loading of devices} and the highest SE in trade-off; on the other hand, \ac{mMIMO} with classical WF achieved the highest SE-EE trade-off; however, the percentage of actives devices is around 0.5.  Fig. \ref{fig:SE-EE-compAll}.(b) depicts results when $M = 128$, NOMA with EPA achieved SE-EE trade-off with the highest {loading of devices}, in contrast, mMIMO with classical WF with lower {loading of devices} achieved higher values of SE and EE in trade-off with half of the active devices. Fig \ref{fig:SE-EE-compAll}.(c) showed the results when M=256, NOMA with $\Delta$-WF achieved SE-EE trade-off with the highest {loading of devices}, and one more time mMIMO with WF achieved higher SE-EE trade-off {with 47\% of active devices}. It is possible to demonstrate that the increase of antennas in the BS improves the SE result, on the other hand, it worsens the EE result. 

\begin{table}[!htbp]
\centering
\caption{SE-EE Trade-off}
\begin{tabular}{llllll}
\hline
\multicolumn{6}{c}{\textbf{M = 64}}             \\ \hline \hline
\multicolumn{1}{l|}{} & \multicolumn{1}{c|}{$\rho$} & \multicolumn{1}{c|}{SE} & \multicolumn{1}{c|}{EE} & \multicolumn{1}{c|}{Actives Users} & $\mathcal{F}$ \\ \hline
\multicolumn{1}{l|}{mMIMO-WF} & \multicolumn{1}{l|}{0.652} & \multicolumn{1}{l|}{131.15} & \multicolumn{1}{l|}{\textbf{0.762}} & \multicolumn{1}{c|}{.50} & .485 \\ \hline
\multicolumn{1}{l|}{NOMA-EPA} & \multicolumn{1}{l|}{\textbf{0.875}} & \multicolumn{1}{l|}{\textbf{147.45}} & \multicolumn{1}{l|}{0.428} &  \multicolumn{1}{c|}{1.0} & .485 \\ \hline
\multicolumn{1}{l|}{NOMA-$\Delta$-WF} & \multicolumn{1}{l|}{0.844}  & \multicolumn{1}{l|}{143.48} & \multicolumn{1}{l|}{0.441} & \multicolumn{1}{c|}{1.0} & .475 \\ \hline
\multicolumn{6}{c}{\textbf{M = 128}}
\\ \hline \hline
\multicolumn{1}{l|}{} & \multicolumn{1}{l|}{$\rho$} & \multicolumn{1}{c|}{SE} & \multicolumn{1}{c|}{EE} & \multicolumn{1}{l|}{Actives Users} & $\mathcal{F}$  \\ \hline
\multicolumn{1}{l|}{mMIMO-WF} & \multicolumn{1}{l|}{0.625} &  \multicolumn{1}{l|}{\textbf{243.21}} & \multicolumn{1}{l|}{\textbf{0.745}} & \multicolumn{1}{c|}{.50} & .475     \\ \hline
\multicolumn{1}{l|}{NOMA-EPA} & \multicolumn{1}{l|}{\textbf{0.734}} & \multicolumn{1}{l|}{241.54} & \multicolumn{1}{c|}{0.411} &  \multicolumn{1}{c|}{1.0} & .49               \\ \hline
\multicolumn{1}{l|}{NOMA-$\Delta$-WF} & \multicolumn{1}{l|}{0.703}          & \multicolumn{1}{l|}{226.32}          & \multicolumn{1}{c|}{0.405} & \multicolumn{1}{c|}{.98} & .45 \\ \hline
\multicolumn{6}{c}{\textbf{M =256}}  \\ 
\hline \hline
\multicolumn{1}{c|}{}         & \multicolumn{1}{l|}{$\rho$}         & \multicolumn{1}{c|}{SE} & \multicolumn{1}{c|}{EE} & \multicolumn{1}{l|}{Actives Users} & $\mathcal{F}$  \\ \hline
\multicolumn{1}{l|}{mMIMO-WF} & \multicolumn{1}{l|}{0.578}          & \multicolumn{1}{l|}{\textbf{404.84}} & \multicolumn{1}{l|}{\textbf{0.691}} & \multicolumn{1}{c|}{.47} & .43     \\ \hline
\multicolumn{1}{l|}{NOMA-EPA} & \multicolumn{1}{l|}{0.523}          & \multicolumn{1}{l|}{344.70}          & \multicolumn{1}{l|}{0.380} &  \multicolumn{1}{c|}{1.0} & .495            \\ \hline
\multicolumn{1}{l|}{NOMA-$\Delta$-WF} & \multicolumn{1}{l|}{\textbf{0.594}} & \multicolumn{1}{l|}{333.31}          & \multicolumn{1}{l|}{0.375}  & \multicolumn{1}{c|}{.85} & .398             \\ \hline
\end{tabular}
\label{tab:table3}
\end{table}

\section{Conclusion and Future Works} \label{sec:concl}
This work proposes a comparative SE and EE analysis in DL single-cell between mMIMO and NOMA with BS equipped with three antenna configurations. Under the SE perspective, mMIMO with the classical WF algorithm achieved better low- and medium-loading results. On the other hand, when the system loading is higher as $\rho > 0.6$ the NOMA achieves better results in the range $0.6 < \rho \leq 2$.

{The analyzed \ac{PA} methods applied to the NOMA system, (EPA, PICPA, and $\Delta$-WF) result in different SE performance}. Indeed, when the channel hardening condition is fully attained, and the amount of BS antennas increases($M =128$ and $256$), the best SE results {are} attained with the proposed $\Delta$-WF algorithm, but, as expected, the fairness index is harmed.

Under the EE perspective, the mMIMO achieved better results when employing the three EPA, PICPA, and WF \ac{PA} methods under $K < M$. However, the NOMA can operate under higher system loading, {\it i.e.}, $K < 2M-1$.

In terms of area-under-SE-curve and EE-curve metrics, $\mathcal{S}$ and $\mathcal{E}$, respectively, the NOMA system attained better results, due to its ability to serve a larger number of users than mMIMO. Such numerical results confirm NOMA's ability to operate with high {loading of devices}. On the other hand, achieving high fairness with NOMA is impossible. 

From the perspective of SE-EE trade-off, mMIMO achieved the best results, because of the superiority in EE; always achieved in {loading of devices} $\rho =$ 0.6 in all $M$ setups.

NOMA systems present exciting features and have been intensively investigated as a promising technique in devising future wireless generations. As future works, hybrid NOMA systems and alternative techniques such as {\it rate-splitting multiple access} (RSMA) can improve the overall EE of massive MIMO systems.

\section*{Acknowledgement}
This work was partly supported by The National Council for Scientific and Technological Development (CNPq) of Brazil under Grants 310681/2019-7, partly by the CAPES- Brazil - Finance Code 001,  and the Londrina State University - Paraná State Government (UEL).


\end{document}